\documentclass[prd,amsfonts,onecolumn,superscriptaddress,aps,nofootinbib,11pt]{revtex4-1}

\usepackage{amsmath, amsthm, amssymb, amsfonts, mathrsfs}
\usepackage{mathtools}
\usepackage{braket, slashed, bm}
\usepackage{hyperref}
\usepackage{booktabs}
\usepackage{scalefnt}
\usepackage{setspace}
\usepackage{amsmath, amsthm, amssymb, amsfonts, mathrsfs}
\usepackage{scalefnt}
\usepackage{braket, slashed, bm}
\usepackage{bbold}
\usepackage{listings}
\usepackage{siunitx}
\usepackage{booktabs}
\usepackage{soul}
\usepackage{color}
\usepackage[dvipsnames]{xcolor}
\usepackage{graphicx}
\usepackage{float}
\usepackage{placeins}
\usepackage{booktabs}
\usepackage{hyperref}
\usepackage[caption=false]{subfig}
\usepackage{soul} %\st
\usepackage{ulem}
\usepackage{multirow}

\hypersetup{
	unicode=false,          % non-Latin characters in Acrobat’s bookmarks
	pdftoolbar=true,        % show Acrobat’s toolbar?
	pdfmenubar=true,        % show Acrobat’s menu?
	pdffitwindow=false,     % window fit to page when opened
	pdfauthor={William},     % author
	colorlinks=true,       % false: boxed links; true: colored links
	linkcolor=blue,          % color of internal links
	citecolor=magenta,        % color of links to bibliography
	urlcolor=blue
}

\begin{document}
%\setkeys{apsrev4-1}{longbibliography=false}

\title{Topic Modeling in New Physics Detection}

\author{Alexandre Alves}
\email{aalves@unifesp.br}
\affiliation{Departamento de F\'isica, Universidade Federal de S\~ao Paulo, UNIFESP, 09972-270, Diadema-SP, Brazil}
\author{Eduardo da Silva Almeida}
\email{almeidae@ufba.br}
\affiliation{Departamento de Física do Estado Sólido, Universidade Federal da Bahia,\\
UFBA, 40170-115, Salvador-BA, Brazil }
\author{Douglas Roberto Pimentel}
\email{douglas.pimentel@unifesp.br}
\affiliation{Departamento de F\'isica, Universidade Federal de S\~ao Paulo, UNIFESP, 09972-270, Diadema-SP, Brazil}

\date{\today}
	
\begin{abstract}
    In this work, we apply topic modeling to detect new physics in proton-proton collisions at the LHC in an unsupervised way. We investigate three new physics scenarios where fully leptonic $t\bar{t}\to b\bar{b}\ell^+\ell^-\nu_\ell\bar{\nu}_\ell$ is the main source of background without relying on jet substructure variables. We demonstrate that the algorithm remains effective even in this low-particle multiplicity framework, complementing jet tagging studies, where it is typically employed. Moreover, we demonstrate that the performance of topic modeling is competitive or even better than well-known outlier detectors, such as isolation forest and variational autoencoders, with moderate and high background pollution in almost all new physics scenarios considered.
\end{abstract}

\maketitle

\section{Introduction}

With the discovery of the Higgs boson in 2012 at the LHC, the Standard Model (SM) of particle physics filled its last major experimental gap. However, several fundamental questions remain unanswered within this framework, such as the need for dark matter to explain the cohesion of galactic structures, dark energy driving the accelerated expansion of the Universe, and the matter–antimatter asymmetry observed in the cosmos \cite{Kasieczka2021}, among others.

Over the past decades, numerous theoretical frameworks have been proposed to extend the Standard Model and address these open questions. As a result, LHC searches have largely adopted a top-down strategy, in which analyzes are designed to test specific Beyond Standard Model (BSM) approaches by generating simulated signal and background events. While this approach has proven effective for well-motivated scenarios, it inherently narrows the exploration space and may overlook unexpected phenomena.

Although many analyses present their outcomes in a nominally “model-independent” form, the event selection and background estimation procedures are still guided by model-specific assumptions. This limitation highlights the importance of developing truly model-independent techniques that can discover new patterns directly from experimental data~\cite{Kasieczka2021}.

In response to this challenge, the particle physics community has increasingly turned to model-independent search strategies capable of detecting potential signals of new physics without assuming a particular theoretical model. In this context, unsupervised Machine Learning (ML) techniques have proven to be particularly promising and have received growing attention in recent years \cite{Kasieczka2021, Bardhan2024}.

One such technique is Topic Modeling (TM), originally developed in the field of Natural Language Processing (NLP) to extract latent topics from a collection of texts without the need for labeled data~\cite{Churchill2022}. Although originally motivated by NLP problems, Metodiev and Thaler~\cite{Metodiev2018} observed a conceptual correspondence between the generative process of text documents and the formation of hadronic jets in high-energy collisions: each collision event can be viewed as a ``document'', and the observables of the resulting jets as ``words''. Under this analogy, the distributions of quarks and gluons produced in collisions can be interpreted as latent topics. The authors demonstrated a proof of concept using the DEMIX model~\cite{pmlr-v33-blanchard14} to extract quark and gluon fractions in mixed Z+jet and dijet samples. This approach has since been explored in several related works \cite{Komiske2018, Brewer2021, Aad2019, Alvarez2020, LeBlanc2023}.

In the studies by Dillon et al.~\cite{Dillon2019, Dillon2020, Dillon2021, Faroughy2021}, the Latent Dirichlet Allocation (LDA) model was employed to identify signal and background components based on patterns learned from jet substructure in proton–proton collisions. Considering QCD dijet events as background, two physical scenarios were investigated in~\cite{Dillon2019, Dillon2020, Dillon2021, Faroughy2021}: one involving $t\bar{t}$ production as a signal, and another simulating a hypothetical BSM process with a heavy charged vector boson and a scalar field.

In these works, a detailed reconstruction of the jet history was required to compute suitable observables for training the algorithm. However, this dependency on jet-substructure information limits the applicability of such methods to more general searches. At the same time, the field of topic modeling encompasses a broad family of algorithms developed to address different types of data, document sizes, and analytical goals. Despite this diversity, no systematic comparison has yet been performed among distinct topic-modeling approaches applied to the same physical scenario.

In our work, we present, for the first time, an approach to apply topic modeling algorithms for detecting new physics signals in LHC collisions without relying on jet substructure variables and scenarios with low particle multiplicities. Our methodology explores only high-level final-state particle observables, significantly simplifying data preparation. Furthermore, we compare the performance of three distinct topic modeling algorithms (LDA \cite{Blei2003}, Biterm (BTM)~\cite{Yan2013}, and ProdLDA \cite{Srivastava2017}) and discuss their respective strengths and limitations within the context of particle physics applications.

We focus our studies on three new physics scenarios that have totally leptonic $t\bar{t}$ backgrounds in common: (a) a heavy new Higgs boson decaying to SM Higgs bosons, $hh$, and also in $t\bar{t}$, (b) a non-resonant $hh$ production with non-SM trilinear coupling, and (c) effective operators affecting top-quark pairs production. We will show that resonant production of Higgs and top-quark pairs can be efficiently identified even in scenarios with high background pollution, while non-resonant $hh$ and effective operators are more challenging scenarios.

Performance comparisons are also made against well-known outlier detector models. Topic modeling is competitive and, in some cases, even superior to isolation forest (IF) \cite{liu2008isolation} and variational autoencoder (VAE) \cite{nguyen2024variational}, particularly in the most challenging non-resonant scenarios.

The manuscript is organized in this way. In section~\ref{section:topic_modeling} we describe the topic modeling algorithms; new physics models targeted by the algorithms are presented in section~\ref{section:new-physics}; details of simulations can be found in section~\ref{section:simulations}; in section~\ref{section:training} we describe the training procedure of the algorithms; section~\ref{section:results} presents the results of the work, while sections~\ref{section:discussions} and \ref{section:conclusions}, the discussions and conclusions, respectively.

\section{Topic Modeling}
\label{section:topic_modeling}

%- Topic modeling is an unsupervised technique in machine learning in which the task is to extract the topics presented in a set of documents (corpus of texts).

%- Identifying these topics is used for many proposals, such as identifying topics and sentiments on social networks \cite{Krestel2009}, as in recommendation systems and information retrieval \cite{Yi2008}.

%- It is important to mention that, due to many different types of text (for example, long texts as scientific articles are so different than social media tweets), there is no one topic modeling algorithm suited for all the cases.

%- Many models have been developed since 1990 \cite{Deerwester1990}, when the first topic modeling algorithm (LSI) was introduced, each suitable for a specific kind of document \cite{Churchill2022}.

Topic modeling refers to a class of unsupervised methods whose goal is to identify latent distributions (commonly called topics) present in a collection of documents. These techniques are widely used in text-mining tasks such as theme and sentiment detection in social media, information retrieval, and recommendation systems \cite{Krestel2009, Yi2008}.

A key aspect of topic modeling is that no single algorithm is universally optimal: long, highly structured documents, short, noisy texts, and sparse messages each require different modeling strategies. Since the introduction of Latent Semantic Indexing (LSI) in the 1990s \cite{Deerwester1990}, many algorithms have been developed, each tailored to specific data characteristics \cite{Churchill2022}.

In particle physics, topic-based approaches have been explored at the jet level \cite{Dillon2019, Metodiev2018}, where the distribution of constituents inside a jet can be interpreted as a “document”, enabling unsupervised separation of signal and background in LHC events. In this work, we extend this idea to events with final-state particles comprising jets, charged leptons, and recovered neutrino solutions in processes with leptonic $t\bar{t}$ as the main background source, without relying on the jet substructure. We investigate whether latent structures extracted directly from basic observables such as invariant masses can provide physically meaningful discrimination power, even under severe background contamination.

The models approached here belong to the class of probabilistic generative models, where documents are assumed to be generated through latent variables, typically discrete topics (as in LDA and BTM), or, in more recent neural approaches such as ProdLDA, continuous latent representations. All models operate under the bag-of-words assumption, treating word occurrences as exchangeable within each document, which allows ignoring word order while modeling word counts. This exchangeability is formally justified by de Finetti’s theorem~\cite{Hewitt1955SymmetricMO}, which states that any infinite sequence of exchangeable observations can be represented as a mixture model involving latent variables. In topic modeling, this means that word tokens in a document can be interpreted as conditionally independent given hidden structures such as topic proportions or latent vectors.

Once the generative process is defined, the objective is to characterize the posterior distribution, which gives the probability of the latent variable, $z$, conditioned on the observed data,

\begin{equation}
    p(z|w) = \frac{p(w,z)}{p(w)}.
\end{equation}

The core difficulty arises because the denominator, known as \textit{evidence} $p(w)$, which is obtained by marginalizing all hidden variables, is generally analytically intractable to compute. As a result, training these models requires approximating the posterior through numerical or variational inference methods rather than computing it exactly.

\subsection{LDA Model}

Latent Dirichlet Allocation is a generative probabilistic model in which each document in a corpus is represented as a finite mixture over latent topics, denoted by the topic–proportion vector $\vec{\theta}$, and each topic is represented as a probability distribution over the vocabulary, denoted by $\vec{\phi}_k$ \cite{Blei2003}. For each word in position $n$ in a document, a latent topic variable $z_{dn}$ is selected from the document-specific mixture $\vec{\theta}_d$, and then the observed word $w_{dn}$ is drawn from the corresponding topic-word distribution $\vec{\phi}_{z_{dn}}$. The joint distribution of all hidden and observed variables for a document with $N_d$ words is

\begin{equation}
 p(\vec{\theta}_d,\vec{z}_d,\vec{\omega}_d| \vec{\alpha}, \vec{\phi}) = p(\vec{\theta}_d|\vec{\alpha})\prod_{n=1}^{N_d} p(z_{dn}|\vec{\theta}_d)p(\omega_{dn}|z_{dn},\vec\phi),
\end{equation}
where $p(\vec{\theta}_d|\vec{\alpha})$ is the distribution of topics in a given document, which follows a Dirichlet distribution. The parameter $\vec{\alpha}$ controls the mixture of topics: high values of $\vec{\alpha}$ correspond to a document containing a broad mix of topics, while low values of $\vec{\alpha}$ result in a document concentrated on a few topics, and it is a hyperparameter of our model. $p(z_{dn}| \vec{\theta}_d)$ is the probability of observing a topic $z_{dn}$ given the topic distribution $\vec{\theta}_d$. Thus, the likelihood of observing a topic $z_{dn}$ depends on how the topics are distributed according to $p(\vec{\theta}_d|\vec{\alpha})$. On the other hand, $p(\omega_{dn}| z_{dn}, \vec{\phi})$ is the probability of observing a word $\omega_{dn}$ given a specific topic $z_{dn}$. The topic-word distributions $\vec{\phi} = \{\phi_k\}_{k=1}^{K}$ are drawn independently from a Dirichlet prior $\Vec{\phi_{k}} \sim \text{Dir}(\vec{\beta})$. The hyperparameter $\vec{\beta}$ controls how concentrated each topic is over the vocabulary. Since the choice of each topic for each word and each word itself is treated independently, the joint probability density function is the product of all these probabilities. Considering all documents in the corpus, we get
\begin{equation}
    p(\vec{w}, \vec{\theta}, \vec{z}, \vec{\phi}|\vec{\alpha}, \vec{\beta}) = \left[\prod_{k=1}^{K}p(\vec{\phi_{k}}|\vec{\beta})\right] \prod_{d=1}^{D} p(\vec{\theta}_d,\vec{z}_d,\vec{\omega}_d| \vec{\alpha}, \vec{\phi}).
\end{equation}
The variables $\vec{\phi}$, $\vec{\theta}$ and $\vec{z}$ are latent and unobserved. By integrating over $\vec{\phi}$, $\vec{\theta}$, and summing over $\vec{z}$, we obtain the marginal distribution, which represents the probability of generating a document, given by

\begin{equation}
p(\vec{w}|\vec{\alpha}, \vec{\beta}) = \int\int\left[\prod_{k=1}^{K}p(\vec{\phi_{k}}|\vec{\beta})\right]\left[ \prod_{d=1}^{D} p(\vec{\theta}_d|\vec{\alpha})\right]\prod_{n=1}^{N_d} \sum_{z_{dn}=1}^K p(z_{dn}|\vec{\theta}_d)p(\omega_{dn}|z_{dn}, \vec{\phi})d{\vec{\theta}_d} d\vec{\phi}.
\end{equation}
Integrating the result of the sum over all possible topic assignments is analytically intractable. For this reason, LDA is trained using variational inference \cite{Blei2003} or collapsed Gibbs sampling \cite{griffiths2004finding}.

%\textcolor{red}{In the case of jet physics, the correspondence between words, documents, and the corpus is as follows: the bins represent words, a jet represents a document, and the corpus is a combination of jets}. In this case, the topics correspond to the parton-level identity of the jet, for example, whether it originates from a quark or a gluon. The Bag-of-Words assumption, which neglects the clustering order information in jets, prevents the probabilistic model from being used as a reliable jet generator, but it can still be applied for classification purposes.

\subsection{Biterm Model}

The Biterm Topic Model is a topic modeling developed for discovering latent topics in a corpus of short texts, like tweets, SMS, and so on. This is important because traditional models, like LDA, need co-occurrence of words at the document level to capture patterns to define latent topics, which suffer from data sparsity in short texts. The BTM captures corpus-level word co-occurrence by modeling unordered word pairs, called biterm \cite{Yan2013}. Each biterm represents two words that co-occur within a short context, and in the BTM generative process, every biterm is drawn from a particular topic within a mixture of topics.

Similarly to LDA, each topic $z_{n}$ has an associated word distribution $\vec{\phi}_{z_{n}} \sim$ Dir($\vec{\beta}$), where $\vec{\phi}_{z_{n}}$ is a multinomial distribution over the vocabulary. However, now the corpus has a global topic mixture $\vec{\theta}\sim$ Dir($\vec{\alpha}$). Given a biterm $b_n=(w_{n1},w_{n2})$, the probability that it is generated by the model is obtained by marginalizing over the latent topic assignment $z_n$:

\begin{equation}
p(b_n|\vec{\theta}, \vec{\phi}) = \sum_{z_n=1}^{K} p(z_n|\vec{\theta})p(b_n|z_n, \vec{\phi}) = \sum_{z_n=1}^{K}\theta_{z_n} \phi_{{z_n},w_{n1}} \phi_{z_n,w_{n2}}.
\end{equation}
For all biterms in the corpus we have
\begin{equation}
p(\vec{B}|\vec{\theta}, \vec{\phi}) = \prod_{n=1}^{N_B} \left(\sum_{z_n=1}^{K}\theta_{z_n} \phi_{z_n,w_{n1}} \phi_{z_n,w_{n2}}\right).
\end{equation}
The likelihood of the full set of biterms $\vec{B}$ is obtained by integrating out the Dirichlet priors:
\begin{equation}
    p(\vec{B}|\vec{\alpha}, \vec{\beta}) = \int p(\vec{\theta}|\vec{\alpha}) \prod_{k=1}^{K}p(\vec{\phi}_k|\vec{\beta}) \prod_{n=1}^{N_B} \left(\sum_{z_n=1}^{K}\theta_{z_n} \phi_{z_n,w_{n1}} \phi_{z_n,w_{n2}} \right) d\vec{\phi} d\vec{\theta}.
\end{equation}
Just like in LDA model, this integral is intractable, so BTM is trained using approximate inference, usually collapsed Gibbs sampling, where both $\vec{\theta}$ and $\vec{\phi}$ are analytically integrated out and topics for biterms are sampled directly from conditional distributions. Training therefore proceeds by iteratively sampling latent topics and estimating expected sufficient statistics, bypassing the need to compute the true evidence.

Since each biterm is associated with a topic, BTM can capture semantic relationships between words even when they appear in very short or isolated documents. On the other hand, modeling only global biterms does not retain information about which words appear together within each document.
This limits its use when word order or local co-occurrence is important; as a consequence, the BTM is less interpretable at the document level than LDA, which provides the topic distribution per document. However, in practice, a biterm might be better than LDA even for long texts due to semantic relationships.

\subsection{ProdLDA Model}

A big challenge in topic modeling is that any modification in the topic structure typically requires designing a new inference algorithm. Autoencoder Variational Inference for Topic Models (AVITM) was developed to generalize inference in topic models \cite{Srivastava2017}. ProdLDA arises as a concrete application of this framework.

While LDA assumes that words in a document are generated from a mixture of topics,
\begin{equation}
    p(w_{dn}=v|\vec{\theta}_d, \vec{\phi}) = \sum_{k=1}^{K} {\theta_{dk}} \phi_{k, v},
\end{equation}
where the mixture is performed directly in the space of probability vectors, ProdLDA departs from this formulation by performing the combination of topics in the space of natural parameters of the multinomial distribution. Each topic is parameterized by an unconstrained vector $\vec{\beta}_k \in \mathbb{R}^V$, with $\vec{\phi}_k=\text{softmax}(\vec{\beta}_k)$. For a given document, the natural parameters are linearly combined as $\sum_k \theta_{dk}\vec{\beta}_k$, and the word likelihood is defined as
\begin{equation}
    p(w_{dn}=v|\vec{\theta}_d, \vec{\phi}) = \frac{\prod_{k=1}^{K} \phi_{kv}^{\theta_{dk}}}{\sum_{v'=1}^{V}\prod_{k=1}^{K} \phi_{kv'}^{\theta_{dk}}}
\end{equation}
This formulation corresponds to a Product of Experts (PoE) model and typically yields sharper
and more distinct topics.

Unlike LDA, ProdLDA does not place a Dirichlet prior over $\vec{\theta}_d$. Instead, AVITM introduces a continuous latent variable $\vec{z}_d \sim N(\vec{\mu}_d,\vec{\Sigma}_d)$ and defines $\vec{\theta}_d=\text{softmax}(\vec{z}_d)$. The marginalized likelihood of the corpus is obtained by
\begin{equation}
    p(\vec{w}|\vec{\phi}) = \prod_{d}^{D}p(\vec{w}_d|\vec{\phi}),
\end{equation}
where each document likelihood is given by
\begin{equation}
    p(\vec{w}_d|\vec{\phi}) = \int N(\vec{z}_d| \vec{\mu}_d,\vec{\Sigma}_d) \prod_{n=1}^{N_d} p(w_{dn}|\vec{\theta}_d,\vec{\phi})d\vec{z}_d,
\end{equation}
which is again intractable, now due to the nonlinear softmax transformation and the Product of Experts likelihood inside the integral. Therefore, ProdLDA is trained using AVITM. A neural encoder $q_{\varphi}(\vec{z}_d|\vec{w}_d)$ approximates the posterior, while the decoder implements the Product of Experts likelihood. Training maximizes the Evidence Lower Bound (ELBO):

\begin{equation}
    \mathcal{L} = \mathbb{E}_{q_{\varphi}(\vec{z}_d|\vec{w}_d)}[\log p(\vec{w}_d|\vec{\theta}_d,\vec{\phi})] - KL(q_{\varphi}(\vec{z}_d|\vec{w}_d)||p(z)),
\end{equation}
where KL is  Kullbach-Liebler divergence \cite{KLdiv}. We now proceed to the details of the models, simulations of the collision events, and the representation of the data.

\section{New Physics Models}
\label{section:new-physics}
In the next sections, we will test the performance of topic models to separate signals from fully leptonic $t\bar{t}$ backgrounds in an unsupervised way. Three interesting new physics scenarios might arise in this context: a heavy neutral scalar that can decay into (1) SM Higgses and (2) to top quarks, and mass dimension-six effective operators affecting the top quark production. We now present models that fit these phenomenological scenarios.

\subsection{Heavy Scalar}
 Consider a new real scalar singlet under the SM group, $S$. This model is also known as the xSM~\cite{Profumo:2007wc}, and exhibits first-order electroweak phase transitions that can be detected, in principle, via gravitational waves in the next generation of satellite experiments like LISA~\cite{amaroseoane2017laserinterferometerspaceantenna}. The interaction Lagrangian of interest for us is
 \begin{eqnarray}
 V(H,S) &=& -\mu^2 H H^\dagger + \lambda (H H^\dagger)^2 + \frac{a_1}{2}(H H^\dagger) S+ \frac{a_2}{2}(H H^\dagger) S^2 \nonumber \\
 && + \frac{b_2}{2}S^2+ \frac{b_2}{3}S^3+ \frac{b_4}{4}S^4
 \end{eqnarray}
 where $S=v_s+s$, and $H^T=(G^+,(v_{EW}+h_{SM}+iG_0)/\sqrt{2})$, and $v_s$, and $v_{EW}$ represent the new scalar and the SM Higgs vacuum expectation value with $v_{EW}=246$ GeV, respectively. All the parameters of this Lagrangian are real. The physical scalars of the model can be obtained after EWSB by the rotation~\cite{Profumo:2007wc}
 \begin{equation}
     h = \cos\theta\;h_{SM}+\sin\theta\; s,\;\;\;\; H = -\sin\theta\;h_{SM}+\cos\theta\; s\; .
 \end{equation}

 The new heavy scalar, $H$, now interacts with all the SM spectrum and, in particular, with the SM Higgs bosons and top quarks. 
 %These interactions of SM-type and only the mass of the heavy scalar can change the kinematic distributions of bottom jets, leptons, and neutrinos. 
 Within xSM, the couplings of $H$ with $hh$ and $t\bar{t}$, and the mass of the new scalar, as well, can be considered as independent parameters.

 The mixture of the scalars also changes the SM Higgs self-interactions. In this case, the trilinear self-coupling of the SM Higgs and the coupling $Hhh$ are no longer independent. However, it is possible to show that there exist blind spots in $H\to hh$~\cite{Alves:2020bpi}, that is, the coupling vanishes for certain combinations of the Lagrangian parameters, but with sizable trilinear SM Higgs couplings. On the other hand, for a relatively narrow heavy Higgs, the resonant production can be the relevant contribution after selection cuts. We thus can find scenarios with dominant resonant or non-resonant $hh$ production.
 
 Concerning a new heavy scalar, we will consider, effectively, the following interactions in our studies
 \begin{equation}
     {\cal L}_{int} \supset \kappa\lambda_{SM} h^3+\lambda_{Hhh}Hh^2+g_{tt}Ht\bar{t}\; ,
 \end{equation}
 where $\kappa$ is a coupling modifier, and the SM corresponds to $\kappa=+1$.
 
 All these interactions can give rise to final states with two bottom jets, two charged leptons, and neutrinos in $pp$ collisions. In the case of the new scalar production and decay to $hh$ or $t\bar{t}$, we should expect hard yields compared to backgrounds. The new scalar interactions are of SM-type, and only the mass of the heavy scalar can change the kinematic distributions of bottom jets, leptons, and neutrinos. In the case of the non-resonant double Higgs production, the interplay between the trilinear Higgs diagram and the top quark box diagram determines the kinematic distributions of the final state particles, that is, $\kappa$ is the parameter of control. We will assume that resonant and non-resonant $hh$ and $H\to t\bar{t}$ will not occur simultaneously.

 \subsection{Effective Operators}

We also study model-independent modifications of the gluon triple gauge coupling and four-fermion interactions, using as illustrative examples, the following dimension-six operators in the Warsaw basis \cite{Grzadkowski_2010},
\vspace{-0.5cm}
\begin{align}
\mathscr{O}_{\tilde{G}} 
&= f^{ABC}\,\tilde{G}^{A\nu}{}_{\mu}\, G^{B\rho}{}_{\nu}\, G^{C\mu}{}_{\rho}, \\
\mathscr{O}_{qq}^{(1)} 
&= \left(\bar{q}_L \gamma_\mu q_L\right)
   \left(\bar{t}_L \gamma^{\mu} t_L\right),
\end{align}
where $f^{ABC}$ are the structure constants of $\mathrm{SU}(3)_C$, $G^A_{\mu \nu}$ denotes the gluon field strength, $\tilde{G}^A_{\mu \nu}$ its dual, $q_L$($t_L$) the left-handed light-quark (top quark) doublet, and $\gamma^\mu$ the gamma matrices. Note that the gluon operator is CP-odd. While $\mathscr{O}_{\tilde{G}}$ modifies the trilinear gluon coupling in $gg\to t\bar{t}$, $\mathscr{O}_{qq}^{(1)}$ represents a contact interaction involving top quark production in $q\bar{q}$ collisions. At order $1/\Lambda^2$ they contribute via interference with the SM diagrams. We set $\Lambda=1$ TeV and consider only one operator at a time in our simulations.

\section{Simulations and Description of the variables}
\label{section:simulations}
All simulations of signals and $t\bar{t}$ backgrounds in $pp\to b\bar{b}\ell^+\ell^-\nu_\ell\bar{\nu}_\ell,\; \ell=e,\mu$ were performed using \texttt{MadGraph5}~\cite{Alwall:2014hca} at Leading Order QCD. We chose
the NNPDF2.3QED~\cite{Ball:2013hta} for the parton distribution functions. Hadronization and showering of quarks and gluons were performed using \texttt{Pythia 8}~\cite{Sjostrand:2007gs}, while \texttt{Fastjet}~\cite{Cacciari:2011ma} was employed for jet clustering. Detector effects were estimated with \texttt{Delphes3}~\cite{deFavereau:2013fsa} using default setting cards.

The variables that we employ to populate the histograms for each event use combinations of two, three, and four visible particles: the two bottom jets and the two charged leptons, plus reconstructed neutrinos using the Higgsness and Topness method~\cite{Kim:2018cxf}. In this method, the neutrinos' momenta must minimize mass constraints that characterize double Higgs or top pair production. For example, the invariant mass of the two charged leptons and the two reconstructed neutrinos should reconstruct a Higgs boson, while a bottom-jet plus a charged lepton and a reconstructed neutrino compound a top-quark mass. For more details, we refer the reader to the original paper~\cite{Kim:2018cxf}. The quality of the neutrino's momentum solution was investigated in Ref.~\cite{Alves:2025slf} and found to be useful in reconstructing kinematic variables of interest, like the mass of a heavy Higgs boson decaying to two SM Higgs bosons, and other kinematic variables as well. In this work, we demonstrate that these solutions are also beneficial for the unsupervised learning of events with jets, charged leptons, and neutrinos. Discarding the neutrino solutions, however, would leave us with too few particles to create a useful event representation for topic modeling to work properly. Yet, compared to the hundreds of particles available in jets, our data representation solution for this low-particle-multiplicity scenario was successful, as we will show.
 
Taking into account the four visible particles and the two reconstructed neutrinos, we can construct, for each event, all possible subsets of final-state momenta containing $k$ particles, with 
$k = 2,\ldots,6$. For a given value of k the number of such subsets is given by the binomial coefficient 
$C_{6,k} = \binom{6}{k}$. This results in a total of $\sum_{k=2}^{6} C_{6,k}=57$ different variables that will define the size of the document (event), where each entry is drawn from the discrete vocabulary (all possible bins) defined by the full binning of the observables. Most variables we will use for the representation require more than one input 4-momenta, so we discard combinations with only one particle, $k=1$.
Note that for jet topics, the size of the documents tends to be much larger, with hundreds of words (bins), as a jet might be composed of hundreds of particles. Another important difference is that, in our case, the number of particles in the final state is fixed, while for jets, the number of components varies. 
 
Each event, $ev$, is a set of six final state momenta: $ev=\left[p_{b_1},p_{b_2},p_{\ell_1},p_{\ell_2},p_{\nu_1},p_{\nu_2}\right]$ that can be combined into subsets of momenta $\{p\}$ using two up to six momenta to form a variable.

 The kinematic variables used for the search of data representation are:
 \begin{itemize}
     \item Invariant mass: $M(\{p\})=\left( \sum_{p_i\in \{p\}} p^\mu_i \right)^2$
     \item Transverse momentum: $P_T(\{p\}) = P_T(\sum_{p_i\in \{p\}}p_i^\mu)$ 
     %\item Difference in the $\eta\times\phi$ plane: $\Delta R(\{p\})=\Delta R(p_i,p_j\in \{p\})$, in this case, $p_i$ or $p_j$ can be itself the sum of a subset of the events' momenta
 %\end{itemize}
 %and the following jet-inspired ones
 %\begin{itemize}
     \item Transverse momentum fraction: $z(\{p\}) = \frac{\min(P_T(\{p\}))}
     {\sum_{p_i\in\{p\}}P_T(p_i)}$
     \item $\cos\theta^*(\{p\}) = \arctan\left(\frac{\Delta\eta(p_i,p_j\in \{p\})}{2}\right)$
     %\item $\kappa(\{p\}) = z(\{p\})\Delta R(p_i,p_j\in \{p\})$
     \item $k_T(\{p\}) = \min(p_T(p_i),p_T(p_j),p_i,p_j\in\{p\})\Delta R(\{p\})$, where the difference in the $\eta\times\phi$ plane is given by $\Delta R(\{p\})=\Delta R(p_i,p_j\in \{p\})=\sqrt{(\eta_i-\eta_j)^2+(\phi_i-\phi_j)^2}$, in this case, $p_i$ or $p_j$ can be itself the sum of a subset of the events' momenta
     \item $\cos\theta_{WW}(\{p\}) = \frac{\vec{p}_i\cdot\vec{p}_j}{||\vec{p}_i|| ||\vec{p}_j||}$, $p_i,p_j\in\{p\}$, where the 3-momenta are calculated in the rest-frame of the $\ell^+,\ell^-,\nu_\ell,\bar{\nu}_\ell$ system. This is inspired by events containing a Higgs boson decaying to $WW$.
     \item $RIV_t(\{p\}) = \frac{\slashed{M}}{M_{vis}(\{p\})}$, where the invisible mass, $\slashed{M}$, is defined as $\sqrt{p_T(\nu_{t_1})^2+p_T(\nu_{t_2})^2+M^2(\nu_{t_1},\nu_{t_2})}$. Here, $\nu_{t_i}$ represents a reconstructed neutrino from the computation of Topness ordered in $p_T$. The visible mass, $M_{vis}=M(\{p\})$. This variable is shown to improve the prospects of observation of $hh$~\cite{Alves:2025slf} at the LHC. 
 \end{itemize}

\begin{figure}[t!]
    \centering
        \includegraphics[width=0.36\linewidth]{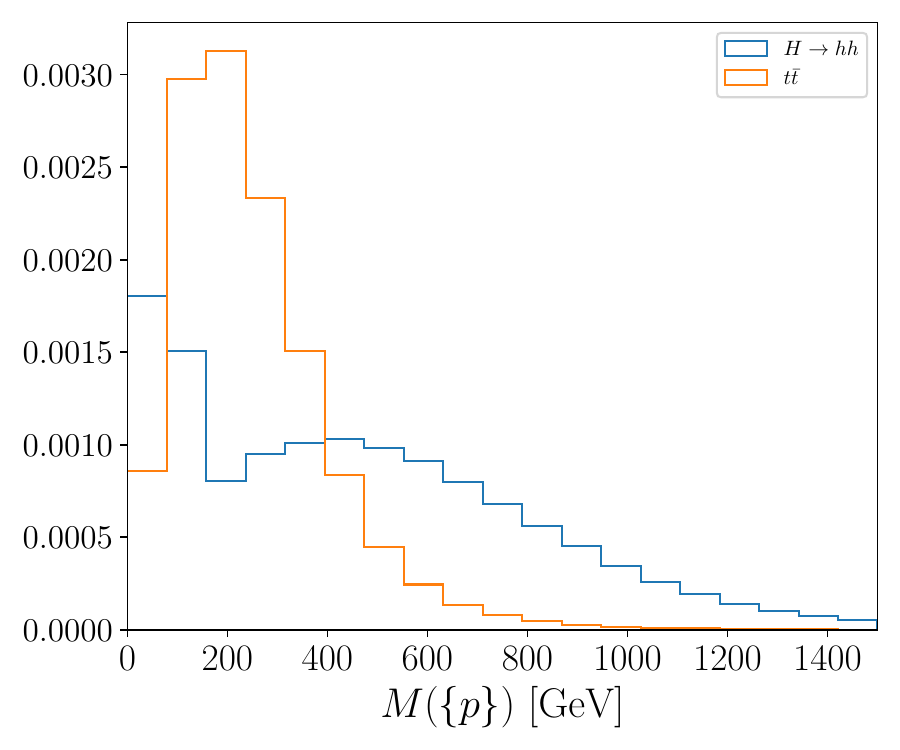}
        \includegraphics[width=0.36\linewidth]{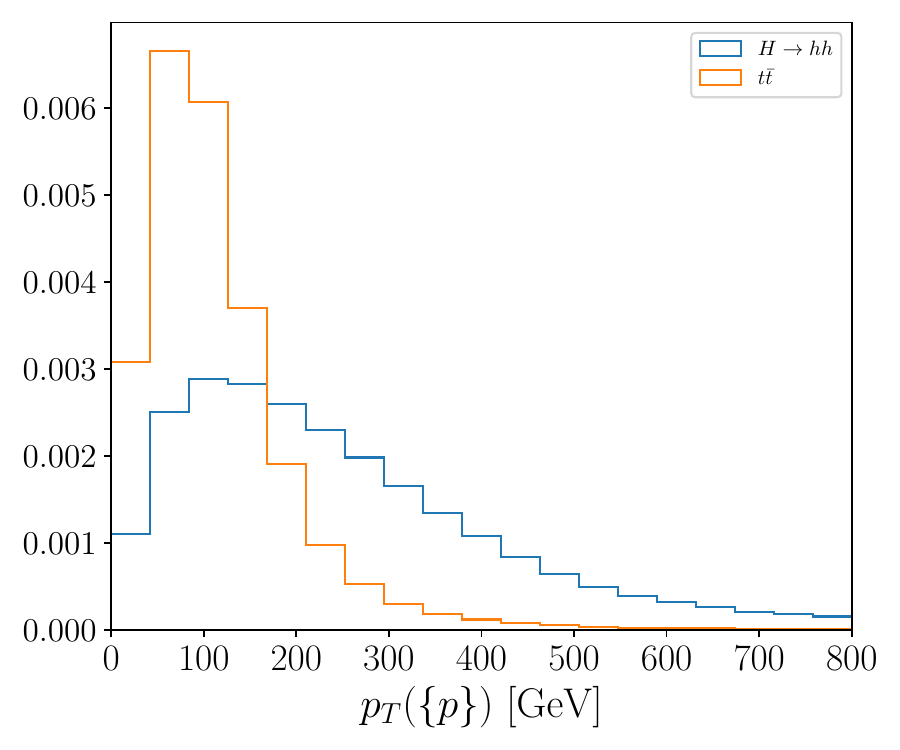}\\
        \includegraphics[width=0.36\linewidth]{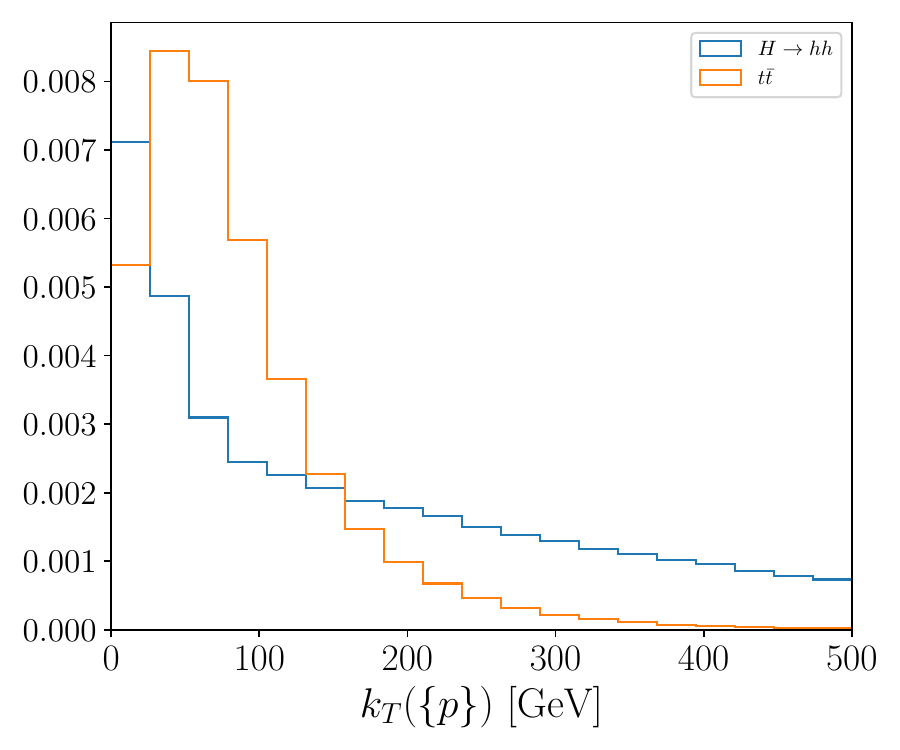}
        \includegraphics[width=0.36\linewidth]{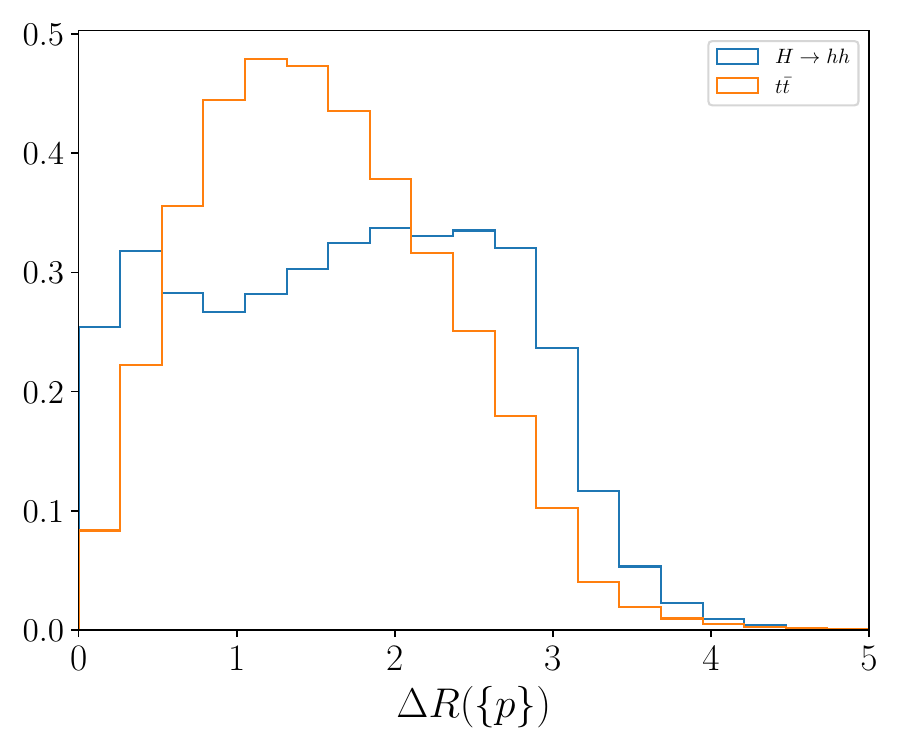}\\
        \includegraphics[width=0.36\linewidth]{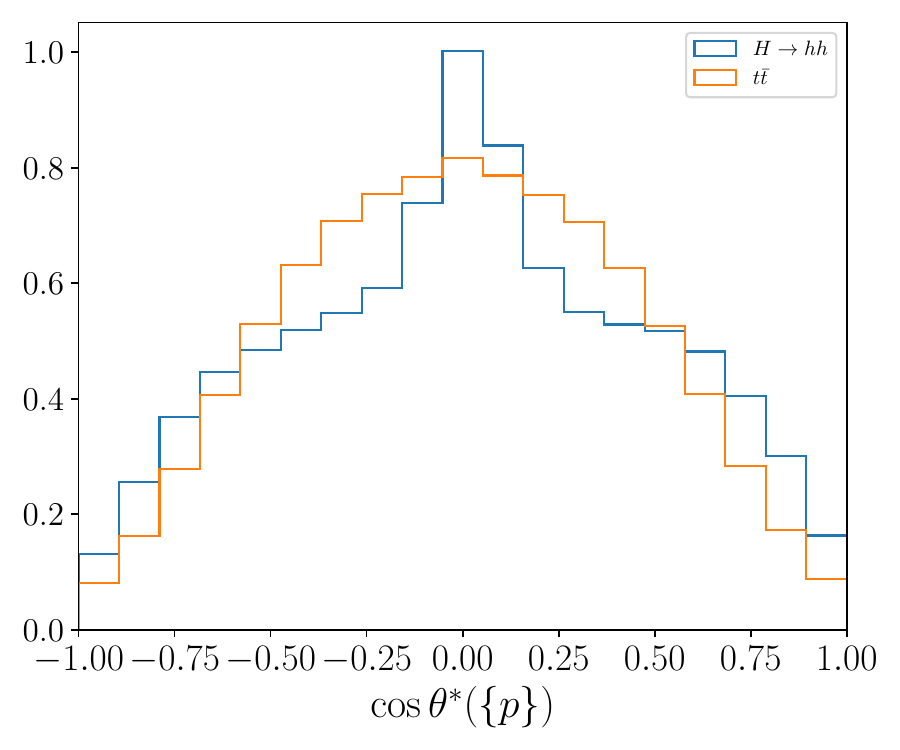}
        \includegraphics[width=0.36\linewidth]{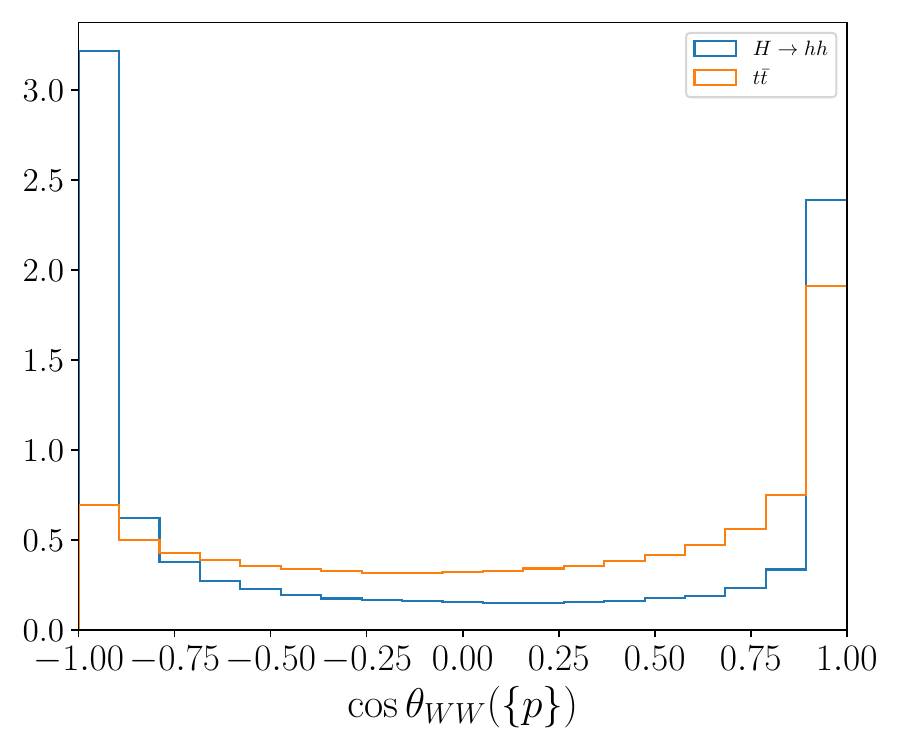}\\
        \includegraphics[width=0.36\linewidth]{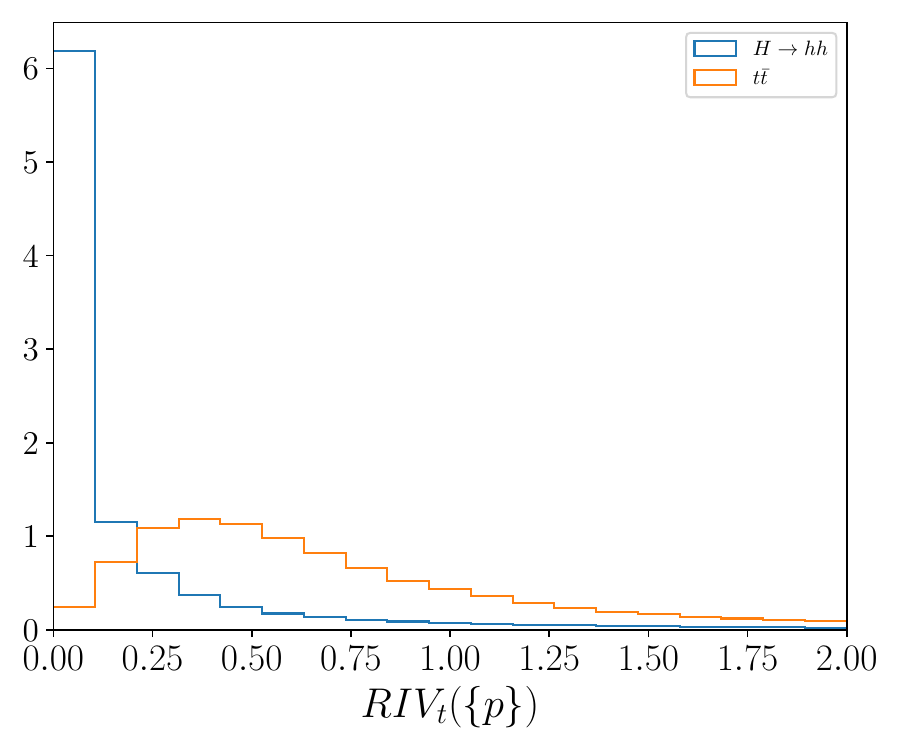}
        \includegraphics[width=0.36\linewidth]{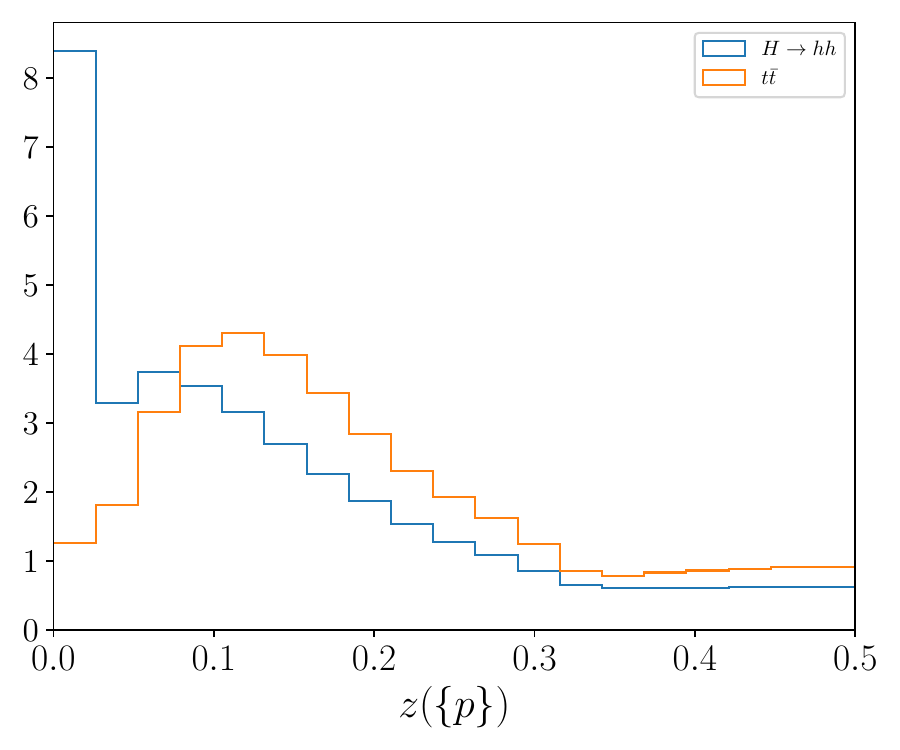}
        \caption{The normalized distributions of the kinematic variables used for the search for the best representation of the data. The signal events correspond to $H\to hh$. The $\Delta R$ variable is used in the computation of $k_T$ but not included explicitly in the representation of the data.}
        \label{fig:dists}
\end{figure}

Now, for each subset $\{p\}\in ev$ we can calculate a 7-D vector $\left[M,p_T,z,\cos\theta_{*},k_T,\cos\theta_{WW},RIV_t\right]$. To form the document that represents the event, we stack all these vectors into fixed-size $57\times N$ arrays. Since each choice of $N$ directly affects the dimensionality of the binning space and, consequently, the size of the vocabulary, it may introduce either additional discriminative information or unnecessary noise.  In practice, not all variables are equally relevant for discrimination, so we searched for the subset of $N$ variables that delivered the best discriminating performance, as we will discuss in the next sections. We display, in Figure~\ref{fig:dists}, the variables used to represent the events plus the $\Delta R(\{p\})$ one, which composes the $k_T$ variable for the case of a 1 TeV Higgs, $\Gamma_H=10$ GeV, decaying to $hh$ and the $t\bar{t}$ background.

It is important to note that the distributions shown in Figure~\ref{fig:dists} take into account all the events of the signal and background samples. The distributions of individual events are much sparser once they are populated by 57 values only. In this respect, the binning of the distributions plays an important role.

An exploratory data analysis was therefore performed to define a binning scheme that spans the full physical range of each observable while preserving the relevant kinematic information and avoiding excessive sparsity at the event level. This discretization enables each event to be represented as a bag-of-words, where each populated bin corresponds to a word, allowing the application of probabilistic topic modeling techniques.

For the observables $M$, $p_T$, and $k_T$, which extend up to approximately 1 TeV, a uniform bin width of 20 GeV was adopted. Angular variables ($\cos\theta^*$ and $\cos\theta_{WW}$) were binned with a width of 0.1 within the interval [-1,1]. The variable $z$, defined in the range [0, 0.5], was discretized using bins of width 0.02, while for $RIV_t$, which extends up to values of order 5, a bin width of 0.2 was used.

After binning, each event is represented by the set of bins populated by its final-state observables. This discrete representation is then used as input to the topic modeling algorithms, treating each event as a document composed of a small number of words drawn from the global vocabulary. Table \ref{tab:tm_to_phys} summarizes the relationship between topic modeling and particle physics features.

\begin{table}[t!]
    \centering
    \renewcommand{\arraystretch}{1.3}
    \begin{tabular}{l|l}
        \hline
        \textbf{Topic Modeling} & \textbf{Particle Physics} \\
        \hline
        Word & Bin in an $N$-dimensional kinematic histogram \\
        Vocabulary & Set of all possible $N$-D bins \\
        Document & 57 bins inside an event, one from each final-state combination ($C_{6,k}$) \\
        Topic & Latent bin distribution (signal/background) \\
        Corpus & Set of all collision events \\
        \hline
    \end{tabular}
    \caption{Correspondence between topic modeling concepts and the representation of final-state particle collision events. The dimensionality $N$ depends on the selected subset of kinematic variables.}
    \label{tab:tm_to_phys}
\end{table}

\section{Training Procedure}
\label{section:training}
Once the representation of each event is defined, we proceed with the application of the topic models. The procedure described in this section for a specific case is directly applicable to all other cases.

We begin by comparing the performance of the three topic modeling algorithms described in Sec. \ref{section:topic_modeling}. All models are trained under the same physical setup, in this case: a 1 TeV Higgs boson decaying into a pair of SM Higgs bosons. We set $\Gamma_H=10$ GeV as the heavy Higgs width.

To identify the most informative set of input variables, we perform a systematic scan in which each model is trained using individual final-state observables. The discriminative power of each configuration is evaluated using the area under the ROC curve. Among the seven available observables, the invariant mass, transverse momentum ($p_T$), and $k_T$ emerge as the most informative variables and are therefore selected for subsequent analyzes.

Biterm Topic Modeling was performed using the \texttt{bitermplus}~\cite{terpilovskii2025bitermplus} package, Latent Dirichlet Allocation using \texttt{gensim}~\cite{rehurek2010gensim}, and ProdLDA was implemented from scratch using the \texttt{pytorch} framework. For each model, a set of algorithm-specific hyperparameters was allowed to vary during training:

\begin{itemize}
    \item Biterm Topic Modeling: $\alpha$ and $\beta$;
    \item Latent Dirichlet Allocation: $\alpha$, $\beta$, number of passes, and number of iterations;
    \item ProdLDA: dropout rate, learning rate, and hidden layer size.
\end{itemize}

In both LDA and BTM, Dirichlet hyperparameters $\alpha$ and $\beta$ can be defined either as scalar (symmetric) or vector-valued (asymmetric) parameters. In this work, we adopt symmetric priors, where $\alpha$ and $\beta$ are single scalar values shared between topics and vocabulary entries, respectively. This choice reduces the number of free parameters and reflects the absence of prior preference for specific topics or observables, which is well-suited for exploratory and physics-driven applications.

For LDA, two additional training parameters are left free: the parameter \texttt{passes} controls the number of full passes over the corpus during model training, while \texttt{iterations} sets the maximum number of internal inference steps used to estimate the document–topic distributions.

For the ProdLDA model, we tune three main hyperparameters: the \texttt{hidden layer size}, the \texttt{learning rate}, and the \texttt{dropout rate}. The size of the hidden layer controls the expressive capacity of the encoder network, determining how complex the latent document representations can be. The learning rate governs the step size of the gradient-based optimization and directly affects convergence speed and training stability. The dropout rate acts as a regularization mechanism, randomly deactivating neurons during training to mitigate overfitting and improve generalization, particularly in regimes with limited or noisy data.

The quality of the models during training was assessed through perplexity minimization, which was used as an intrinsic measure. After training, the AUC was calculated as an extrinsic evaluation metric to quantify the signal–background separation achieved by each configuration.

For the additional physics scenarios considered later in this work, only the Biterm model was employed, given its superior performance in the main benchmark setup. As done here, in each physics scenario approached, we searched for the subset of observables with better discriminative power.

\section{Results}
\label{section:results}

This section presents the quantitative and qualitative results obtained from the proposed approach. As performance metrics, we use the AUC of the ROC curve and the background-rejection versus signal-efficiency curves. In addition, we inspect the reconstructed observable distributions (like invariant mass) to validate the physical consistency of the classification.

For all scenarios, each model is evaluated under two contrasting conditions: a balanced situation with 50\% background pollution and an extremely unbalanced case with 99\% background pollution. For comparison, we contrast the topic modeling results with two commonly used outlier detection models, VAE and Isolation Forest from \texttt{pyOD}~\cite{zhao2019pyod}.

\subsection{Heavy Higgs to SM Higgses}

First, we investigate the detection of Heavy Higgs bosons to $H\to hh$. This process exhibits a distinctive mass distribution, which makes it a suitable benchmark for verifying whether the topic-modeling approach can detect the underlying patterns that distinguish signal from background, and how its behavior changes as the background becomes dominant.

\begin{figure}[t!]
    \centering
    \begin{minipage}{0.48\linewidth}
        \centering
        \includegraphics[width=\linewidth]{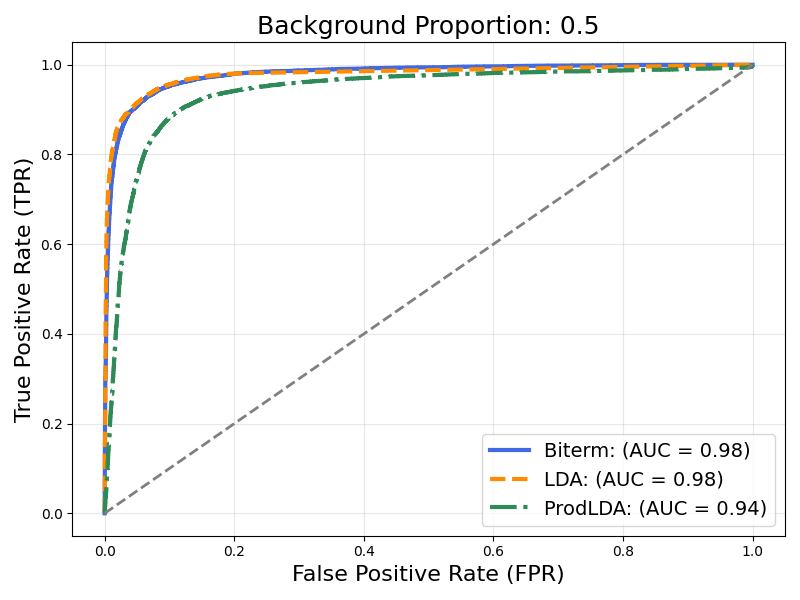}
    \end{minipage}\hfill
    \begin{minipage}{0.48\linewidth}
        \centering
        \includegraphics[width=\linewidth]{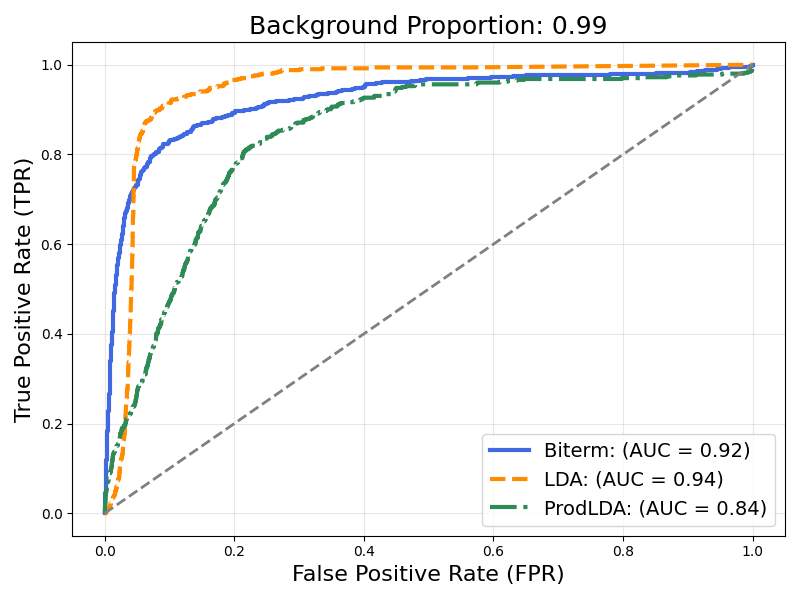}
    \end{minipage}
    \begin{minipage}{0.48\linewidth}
        \centering
        \includegraphics[width=\linewidth]{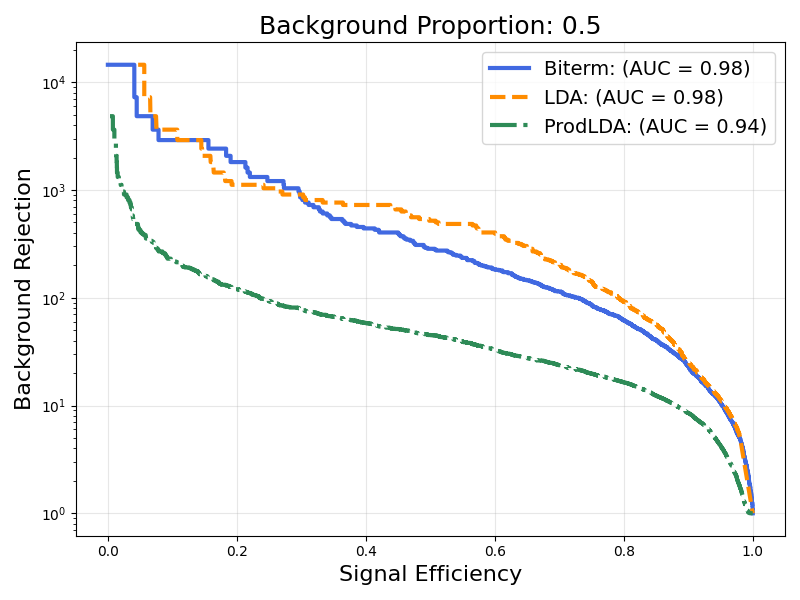}
    \end{minipage}\hfill
    \begin{minipage}{0.48\linewidth}
        \centering
        \includegraphics[width=\linewidth]{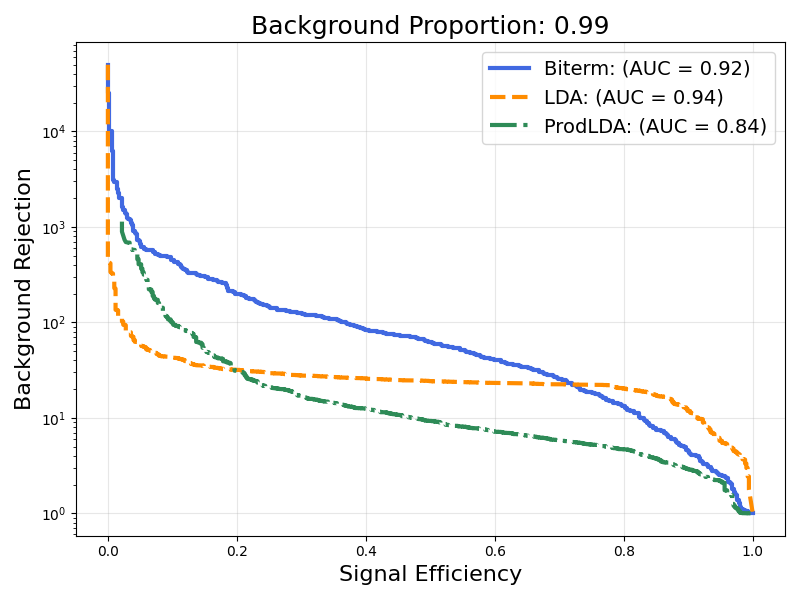}
    \end{minipage}
    \caption{Comparison among the ROC and Background Rejection curves from topic modeling models: LDA, biterm, and ProdLDA. The left column shows the case of 50\% of background pollution, while the right side shows the 99\% scenario. The upper figures show the ROC curves, which represent the relationship between the true positive rate (TPR) and false positive rate (FPR) across many different classification thresholds. The lower figures show the background rejection, which relates the background rejection (inverse of FPR) to the signal efficiency (TPR). The legends show the AUC values, which are used to discriminate the ability of each model in disentangling the signal from the background.}
    \label{fig:roc_curve_Hhh_TM_comparisons}
\end{figure}

Figure \ref{fig:roc_curve_Hhh_TM_comparisons} presents a comparison of the performance of the three topic-modeling approaches described in Section \ref{section:topic_modeling}: LDA, Biterm, and ProdLDA. All models display strong discriminative power between signal and background, with AUC values above 0.9 in all cases, except for ProdLDA in the scenario with 99\% background contamination.

The LDA and Biterm models yield competitive results. While LDA achieves a slightly higher AUC in the 99\% contamination scenario, the Biterm model shows superior background-rejection capability across a broad range of signal efficiencies. Considering its cost–benefit ratio and algorithmic simplicity, we adopt Biterm as the topic modeling method for all subsequent analyses.

\begin{figure}[t!]
    \centering
    \begin{minipage}{0.48\linewidth}
        \centering
        \includegraphics[width=\linewidth]{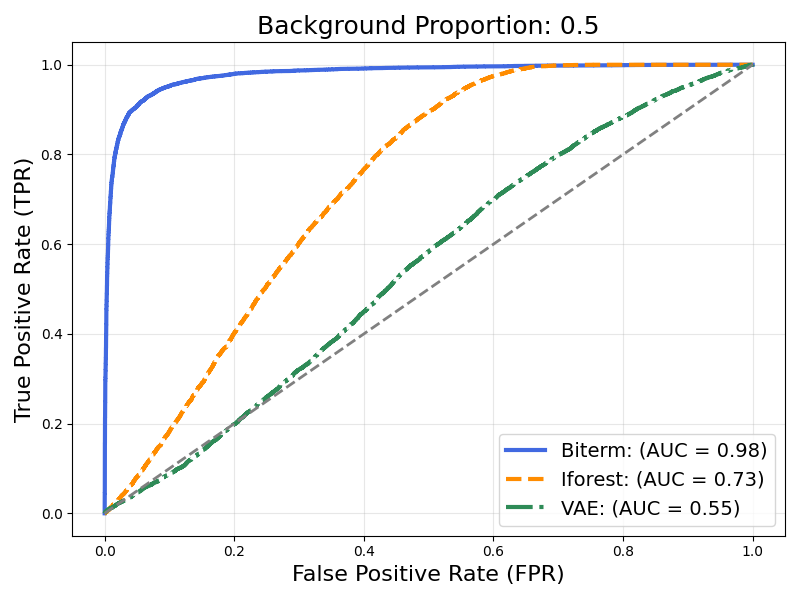}
    \end{minipage}\hfill
    \begin{minipage}{0.48\linewidth}
        \centering
        \includegraphics[width=\linewidth]{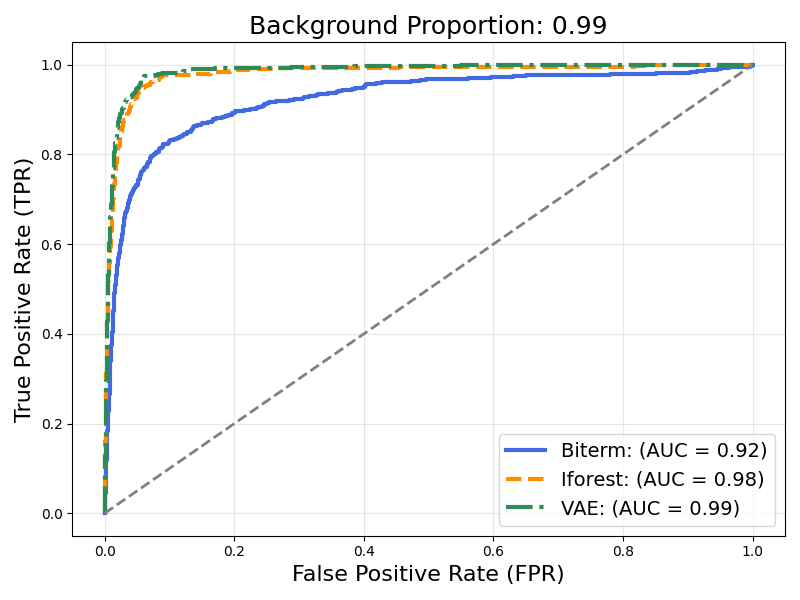}
    \end{minipage}
    \begin{minipage}{0.48\linewidth}
        \centering
        \includegraphics[width=\linewidth]{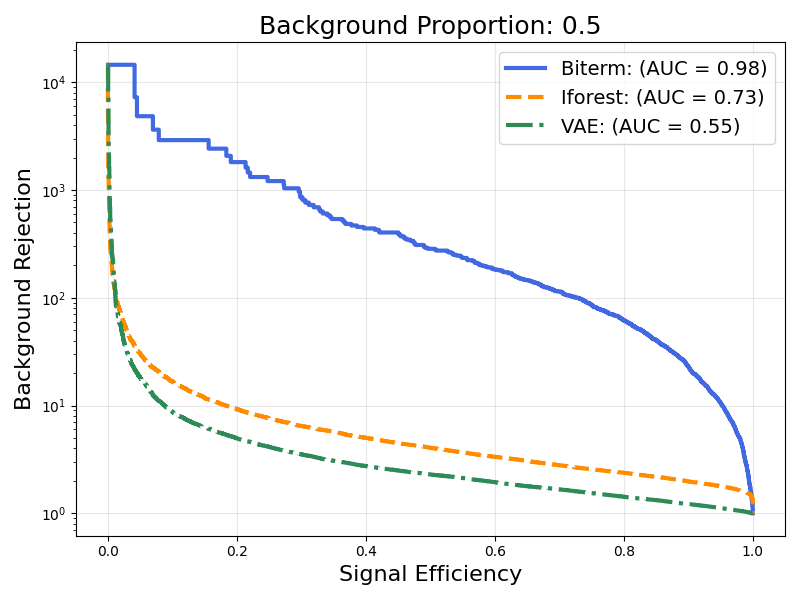}
    \end{minipage}\hfill
    \begin{minipage}{0.48\linewidth}
        \centering
        \includegraphics[width=\linewidth]{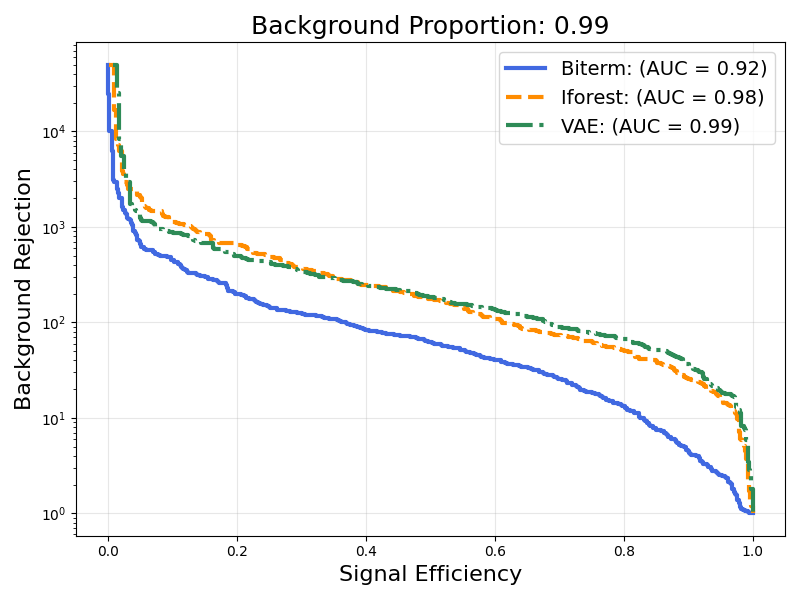}
    \end{minipage}
    \caption{Comparison among the ROC and Background Rejection curves from biterm, isolation forest, and VAE models in $H\xrightarrow{}hh$ process. The left column shows the case of 50\% of background pollution, while the right side shows the 99\% scenario. The upper figures show the ROC curves, which represent the relationship between the true positive rate and false positive rate across many different classification thresholds. The lower figures show the background rejection, which relates the background rejection to the signal efficiency. The legends show the AUC values, which are used to discriminate the ability of each model in disentangling the signal from the background.}
    \label{fig:heavy_higgs_to_dihiggs}
\end{figure}

Figure \ref{fig:heavy_higgs_to_dihiggs} shows the ROC and background-rejection curves for the topic-modeling approach, compared with VAE and Isolation Forest, under the two background-pollution regimes: 50\% (left) and 99\% (right).

In the 50\% background scenario, the Biterm delivers the best overall performance, with AUC = 0.98, and achieves background rejection up to two orders of magnitude higher than VAE and IF at low signal efficiencies (signal efficiency $\approx$ 0.2).

In the 99\% background regime (more favorable to outlier detection methods), VAE (AUC = 0.99) and IF (AUC = 0.98) outperform Biterm, but the topic modeling algorithm still achieves a solid AUC = 0.92 and remains competitive at low signal efficiency, where its rejection power approaches that of the specialized models.

The topic modeling model uses only three final-state observables $\{M, p_{T}, k_{T}\}$, but consistently extracts latent patterns that separate signal from background without relying on jet-substructure information. These results indicate that topic modeling provides a robust unsupervised alternative for signal-background separation and can complement traditional outlier-based approaches.

\subsubsection{Invariant mass reconstruction}

\begin{figure}[htbp]
    \centering
    \begin{minipage}{0.7\linewidth}
        \centering
        \includegraphics[width=\linewidth]{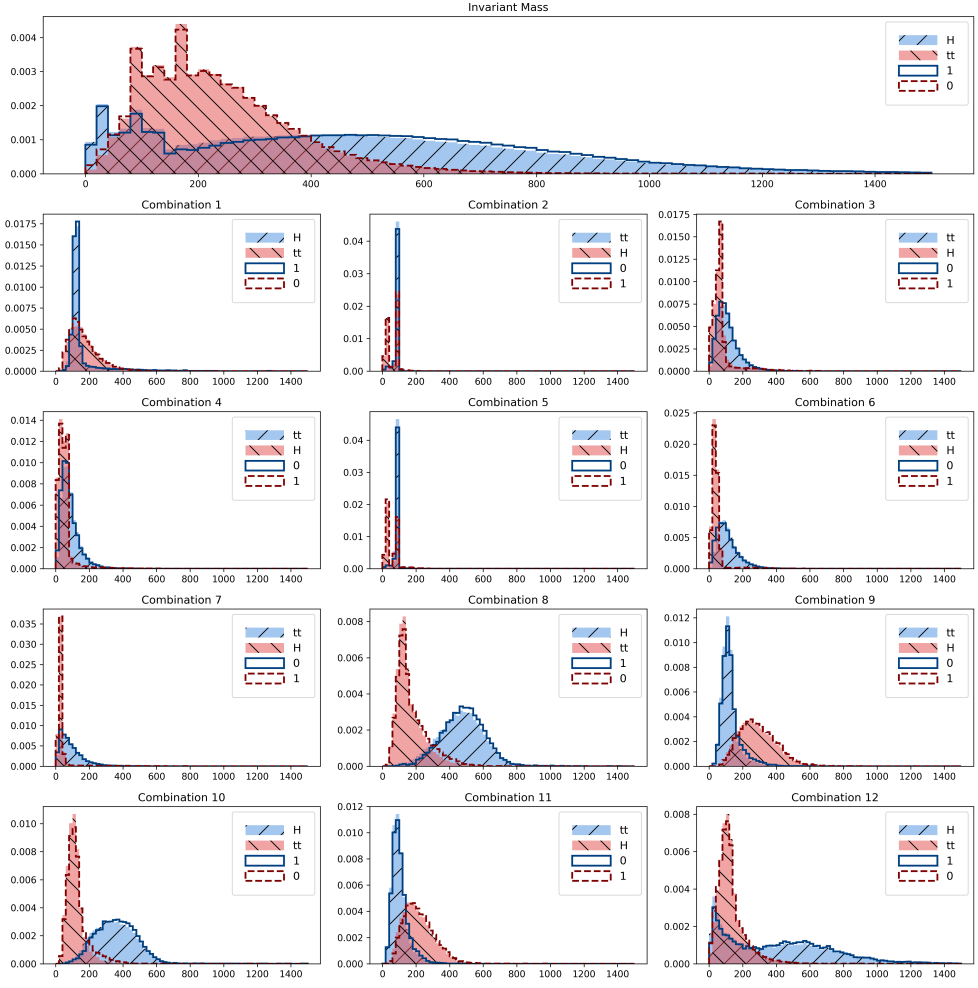}
    \end{minipage}\hfill
    \begin{minipage}{0.7\linewidth}
        \centering
        \includegraphics[width=\linewidth]{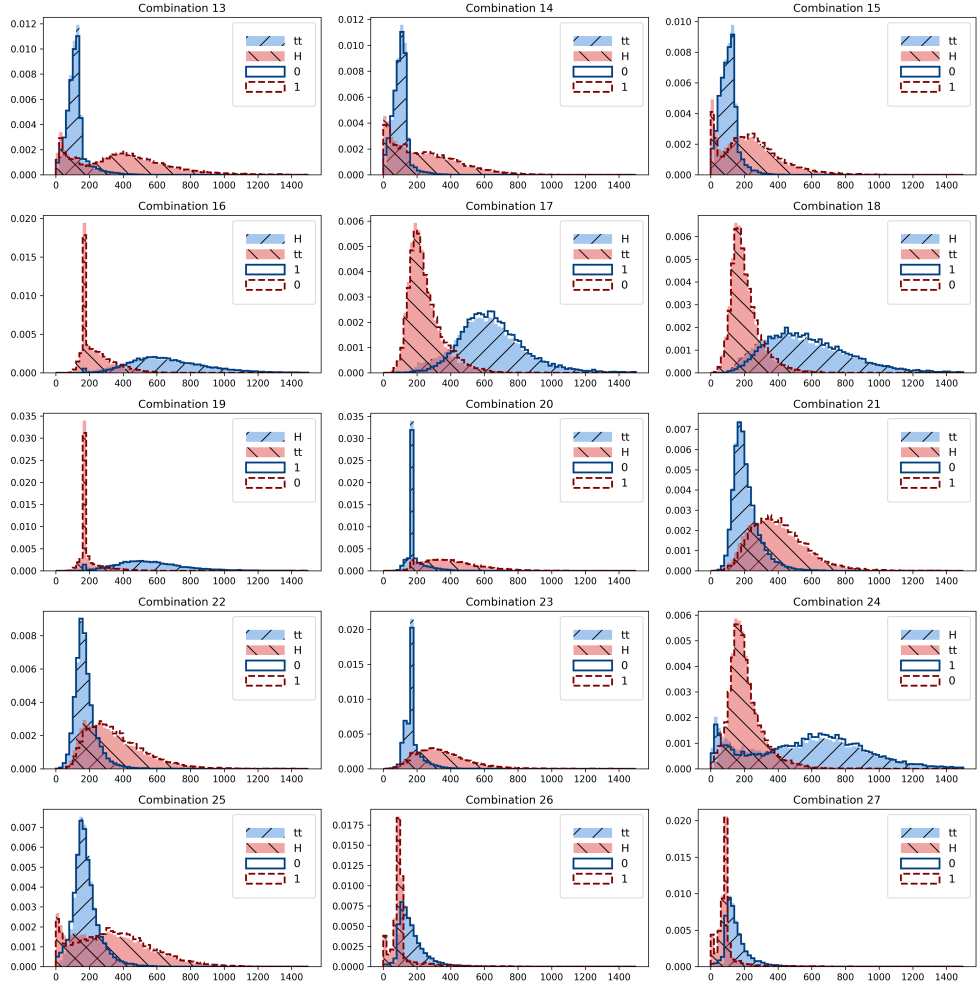}
    \end{minipage}
    \caption{Reconstruction of the invariant mass for $H\xrightarrow{}t\bar{t}$ using BTM with 50\% background pollution. Upper plot shows the distribution of all combinations together. Each mini-plot is the distribution of each possible combination. The filled histograms refers to the event simulated data, while the line histograms refers to the topic classification.}
    \label{fig:mass_reconstructed_Hhh_50_1}
\end{figure}

\begin{figure}[htbp]
    \centering
    \begin{minipage}{0.7\linewidth}
        \centering
        \includegraphics[width=\linewidth]{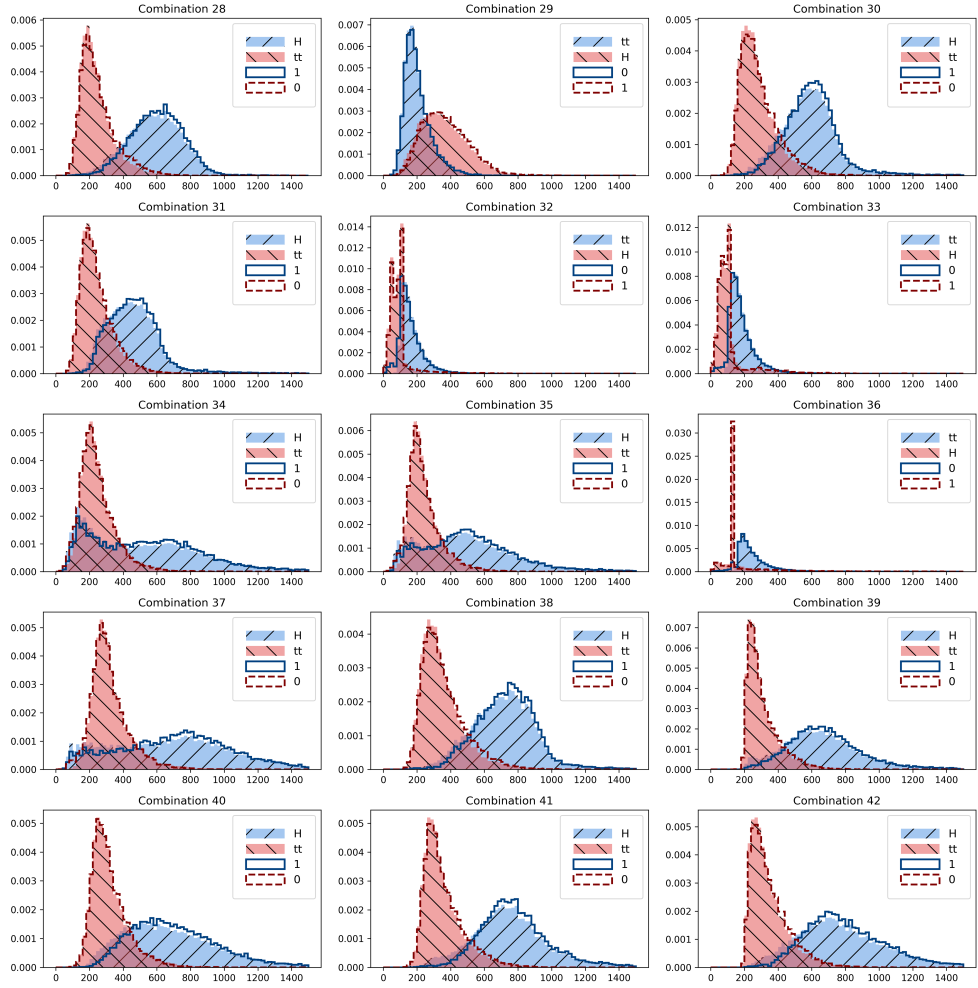}
    \end{minipage}\hfill
    \begin{minipage}{0.7\linewidth}
        \centering
        \includegraphics[width=\linewidth]{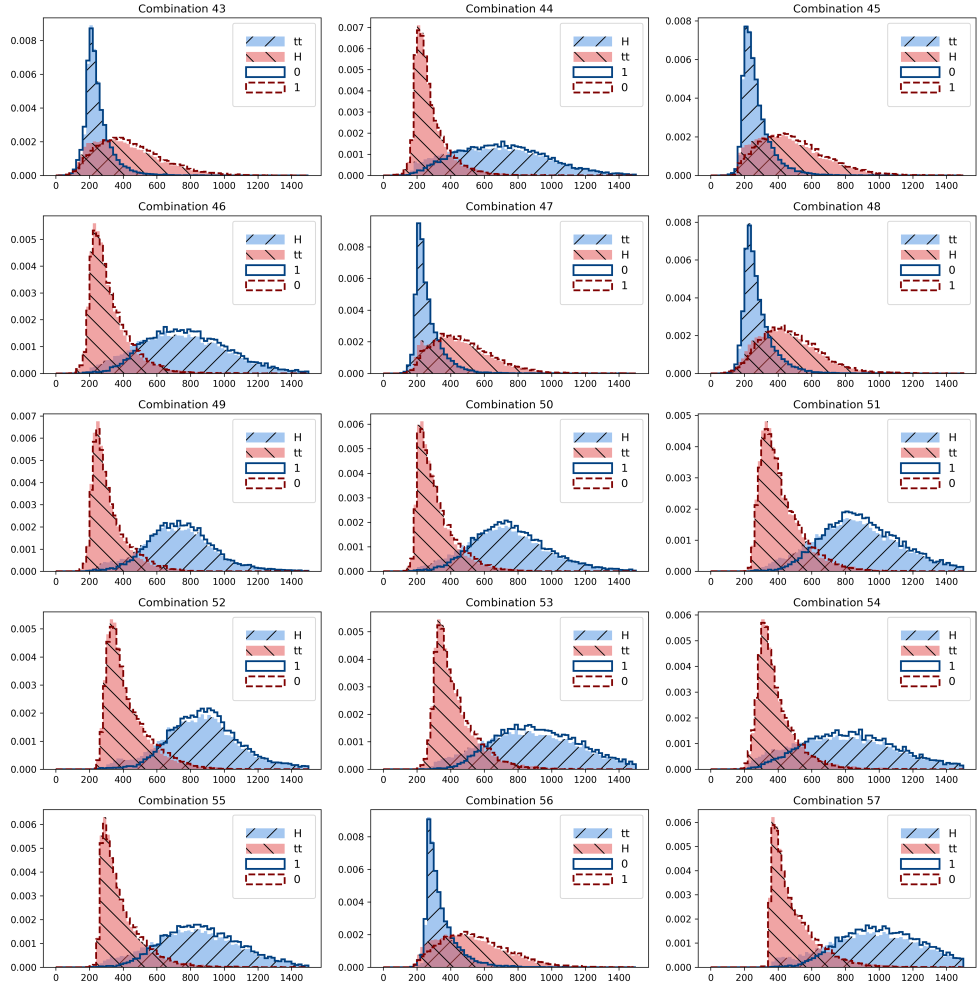}
    \end{minipage}
    \caption{Reconstruction of the invariant mass for $H\xrightarrow{}t\bar{t}$ using BTM for 50\% background contamination. Continuation of the figure \ref{fig:mass_reconstructed_Hhh_50_1}.}
    \label{fig:mass_reconstructed_Hhh_50_2}
\end{figure}

\begin{figure}[htbp]
    \centering
    \begin{minipage}{0.7\linewidth}
        \centering
        \includegraphics[width=\linewidth]{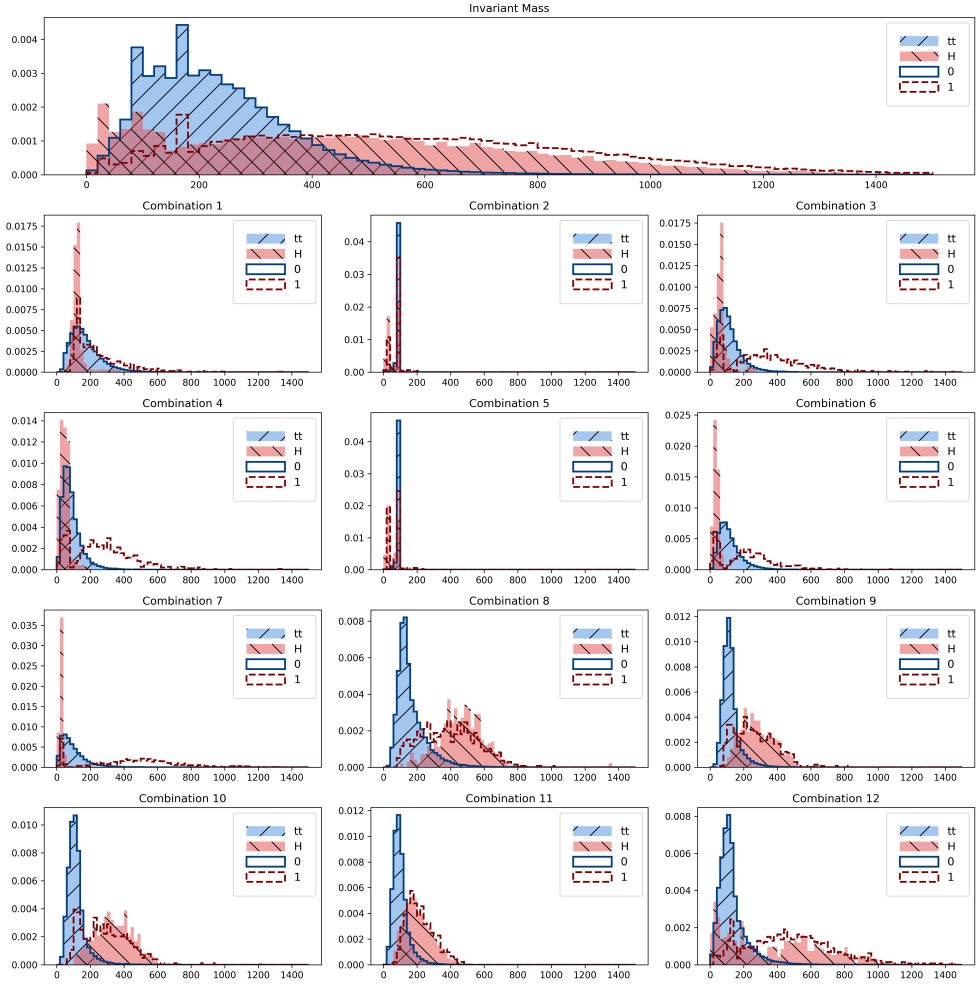}
    \end{minipage}\hfill
    \begin{minipage}{0.7\linewidth}
        \centering
        \includegraphics[width=\linewidth]{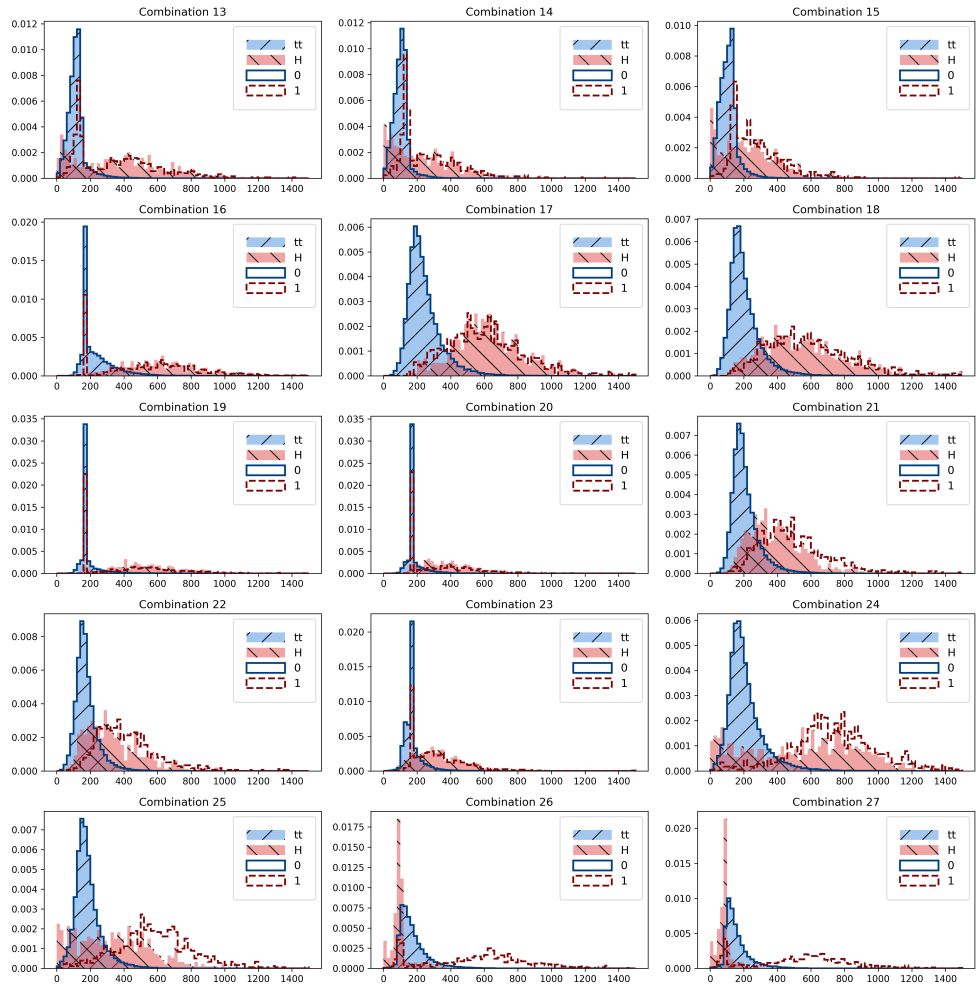}
    \end{minipage}
    \caption{Reconstruction of the invariant mass for $H\xrightarrow{}t\bar{t}$ using BTM with 99\% background pollution. Upper plot shows the distribution of all combinations together. Each mini-plot is the distribution of each possible combination. The filled histograms refers to the event simulated data, while the line histograms refers to the topic classification.}
    \label{fig:mass_reconstructed_Hhh_99_1}
\end{figure}

\begin{figure}[htbp]
    \centering
    \begin{minipage}{0.7\linewidth}
        \centering
        \includegraphics[width=\linewidth]{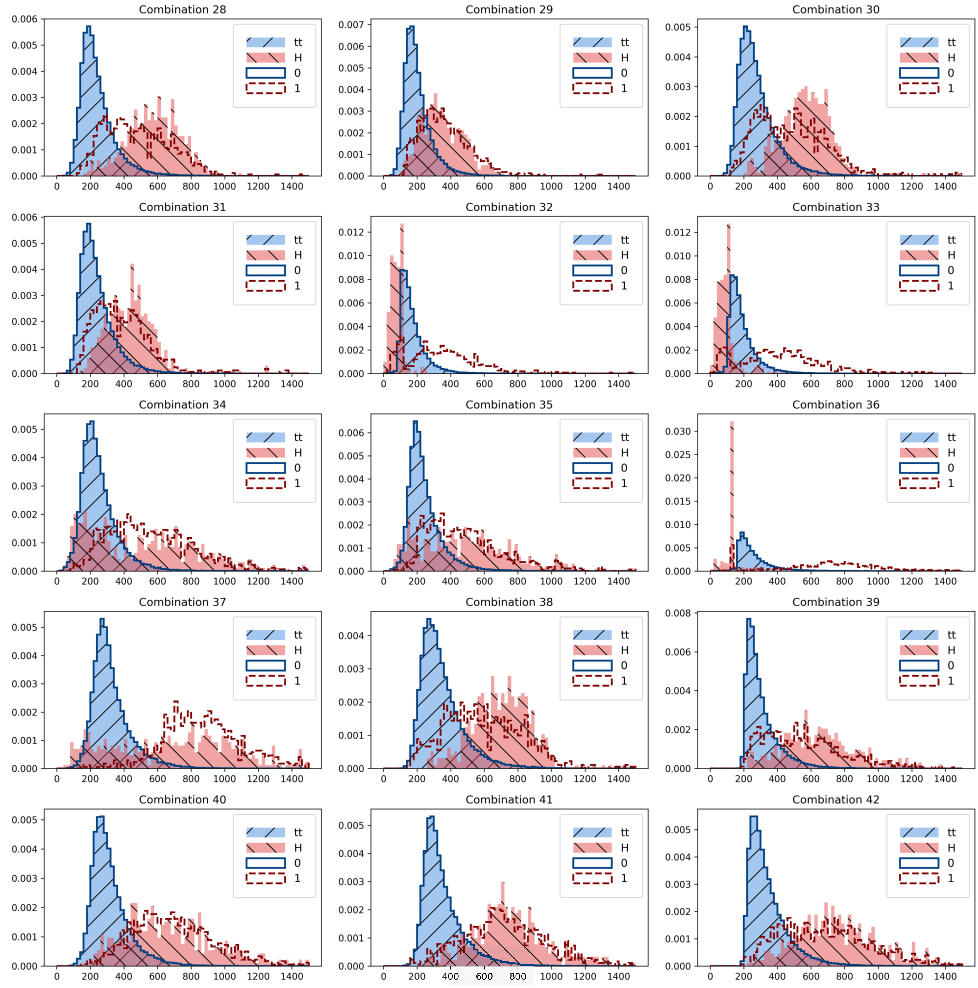}
    \end{minipage}\hfill
    \begin{minipage}{0.7\linewidth}
        \centering
        \includegraphics[width=\linewidth]{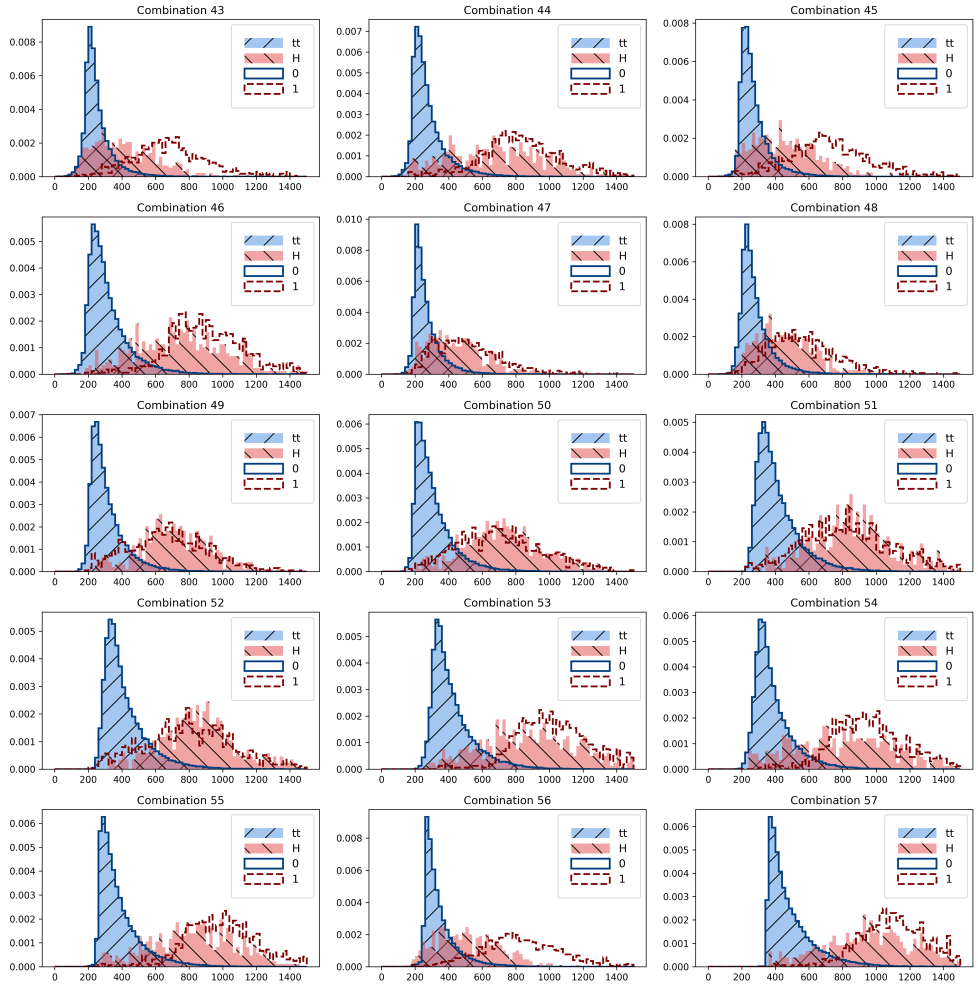}
    \end{minipage}
    \caption{Reconstruction of the invariant mass for $H\xrightarrow{}t\bar{t}$ using BTM for 99\% background contamination. Continuation of the figure \ref{fig:mass_reconstructed_Hhh_99_1}.}
    \label{fig:mass_reconstructed_Hhh_99_2}
\end{figure}

Although ROC and background-rejection curves quantify the statistical discrimination achieved by each model, it is essential to verify whether this separation corresponds to physically meaningful structures. Figures \ref{fig:mass_reconstructed_Hhh_50_1} and \ref{fig:mass_reconstructed_Hhh_50_2} show the reconstructed $M({p})$ distributions for events classified as signal-like and background-like in the 50\% background-contamination scenario; the corresponding plots for the 99\% contamination case are presented in Figures \ref{fig:mass_reconstructed_Hhh_99_1} and \ref{fig:mass_reconstructed_Hhh_99_2}.

In the balanced regime, the reconstructed distributions exhibit a clear agreement with the true ones across all feature combinations used by the topic-modeling approach. Even in the highly polluted 99\% background scenario, the agreement remains good, particularly for background events. Although signal identification naturally becomes more challenging in this regime, the reconstructed tail reproduces the expected physical behavior, indicating that the model still captures the underlying structure of the signal.

A noteworthy example is combination 57, which behaves as an effective estimator of the heavy-Higgs invariant mass. In both background-pollution regimes, this combination yields a reconstructed distribution that closely matches the true mass peak at approximately 1 TeV, reinforcing the physical consistency of the separation learned by the topic modeling model. %\textcolor{red}{Explicar melhor as legendas das figuras 4 e similares}

\subsection{Heavy Higgs to $t\bar{t}$}

The second process approached is the heavy Higgs decaying into a pair of quark and anti-quark top $H\xrightarrow{}t\bar{t}$. Similarly to the $H\xrightarrow{}hh$ case, the set of variables $\{M, p_T,  k_T\}$ provides the best discriminative power between signal and background.

\begin{figure}[t!]
    \centering
    \begin{minipage}{0.48\linewidth}
        \centering
        \includegraphics[width=\linewidth]{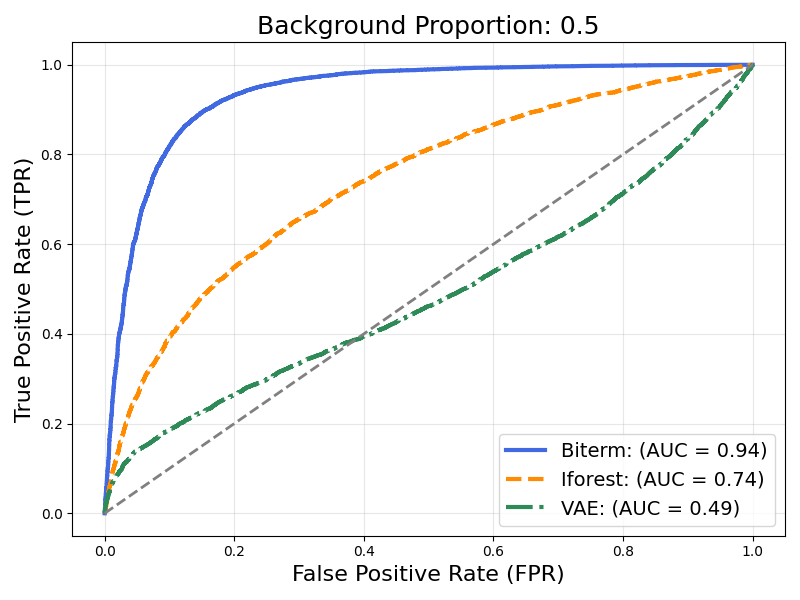}
    \end{minipage}\hfill
    \begin{minipage}{0.48\linewidth}
        \centering
        \includegraphics[width=\linewidth]{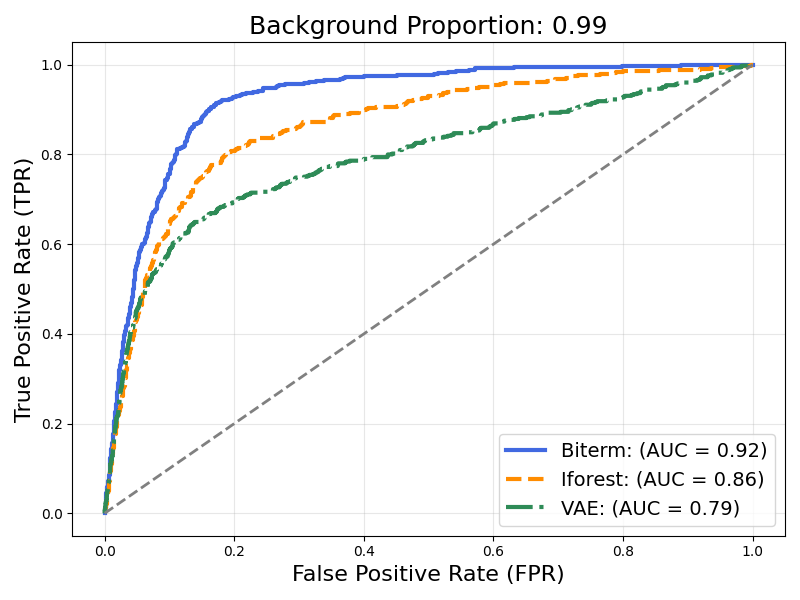}
    \end{minipage}
    \begin{minipage}{0.48\linewidth}
        \centering
        \includegraphics[width=\linewidth]{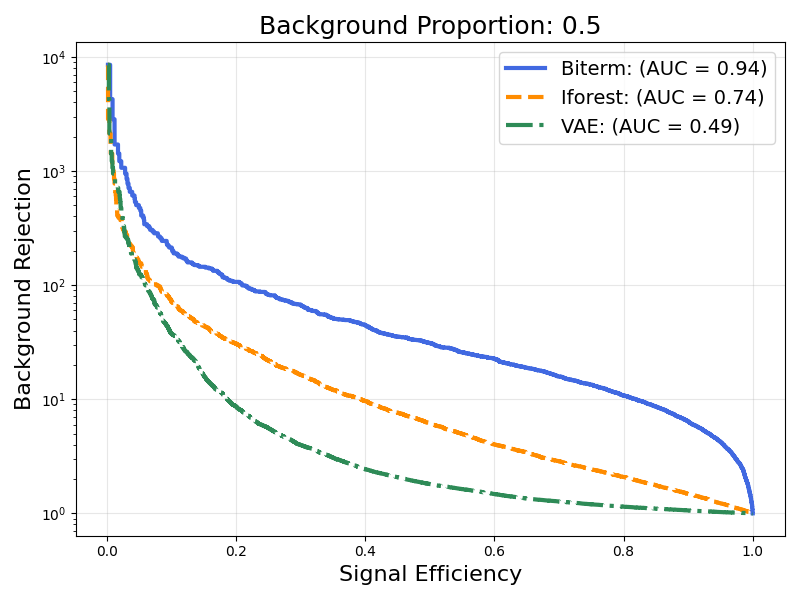}
    \end{minipage}\hfill
    \begin{minipage}{0.48\linewidth}
        \centering
        \includegraphics[width=\linewidth]{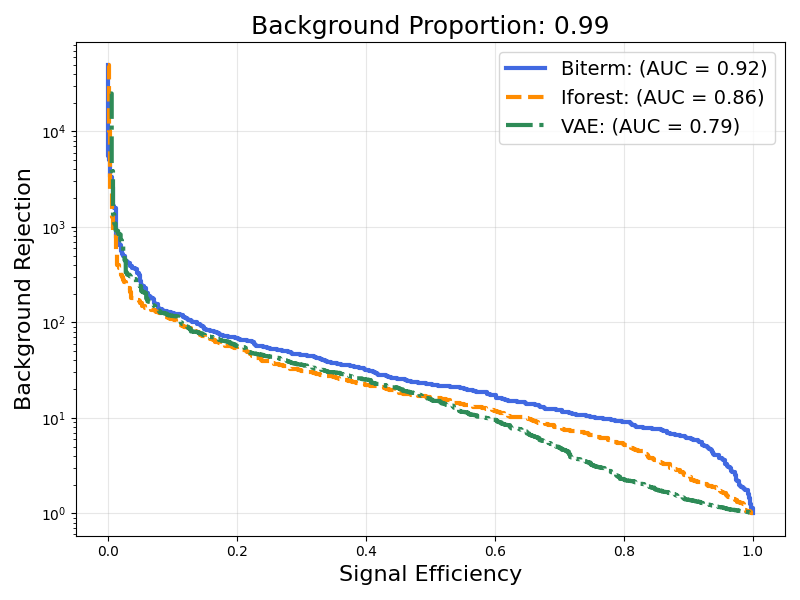}
    \end{minipage}
    \caption{Comparison among the ROC and Background Rejection curves from biterm, isolation forest, and VAE models in $H\xrightarrow{}t\bar{t}$ process. The left column shows the case of 50\% of background pollution, while the right side shows the 99\% scenario. The upper figures show the ROC curves, which represent the relationship between the true positive rate and false positive rate across many different classification thresholds. The lower figures show the background rejection, which relates the background rejection to the signal efficiency. The legends show the AUC values, which are used to discriminate the ability of each model in disentangling the signal from the background.}
    \label{fig:heavy_higgs_to_tt}
\end{figure}

As we can see in Figure \ref{fig:heavy_higgs_to_tt}, in this case, the topic modeling outperforms the alternative models in both background pollution scenarios, with the AUC value higher than $0.9$ in the two cases. The background-rejection curve for $99\%$ pollution shows that the three models are very competitive in terms of rejecting background for a long interval of signal efficiency, with a clearer best performance of the topic model in a high regime of signal detection.

This result reinforces the capacity of topic modeling models to separate background from signal in resonant cases, even in extreme scenarios of high background pollution.

The table \ref{tab:results_heavy_higgs} summarizes the results for the two heavy Higgs scenarios approached, decaying into two Higgses and in a pair $t\bar{t}$.

\begin{table}[h]
\centering
\begin{tabular}{lcc|cc}
\toprule
Model 
& \multicolumn{2}{c}{$H\xrightarrow{}hh$} 
& \multicolumn{2}{c}{$H\xrightarrow{}t\bar{t}$} \\

\cmidrule(lr){2-3}
\cmidrule(lr){4-5}

& 50\% bkg & 99\% bkg & 50\% bkg & 99\% bkg \\
\midrule
BTM     & 0.98 & 0.92 & 0.94 & 0.92 \\
IForest & 0.73 & 0.98 & 0.74 & 0.86 \\
VAE     & 0.55 & 0.99 & 0.49 & 0.79 \\
\bottomrule
\end{tabular}
\caption{
AUC values summarizing the ROC curves for each model and Heavy Higgs production scenarios.
Background contamination levels of 50\% and 99\% were considered.
}
\label{tab:results_heavy_higgs}
\end{table}

\subsection{Non-Resonant Double Higgs Production}

Now we focus on the non-resonant production of a double Higgs, where three possible scenarios are investigated: the SM coupling $\kappa=+1$ and two BSM scenarios which are also not experimentally excluded, $\kappa=-1$ and $\kappa=+3$. The set of variables for $\kappa=+1$ is \{$p_T$, $k_T$, $z$, $\cos\theta_{WW}$\}, for $\kappa=+3$ it is \{$p_T$, $z$, $\cos\theta_{WW}$\} and for $\kappa=-1$ it is \{$p_T$, $k_T$\}.

\begin{figure}[hbpt]
    \centering
    \begin{minipage}{0.48\linewidth}
        \centering
        \includegraphics[width=\linewidth]{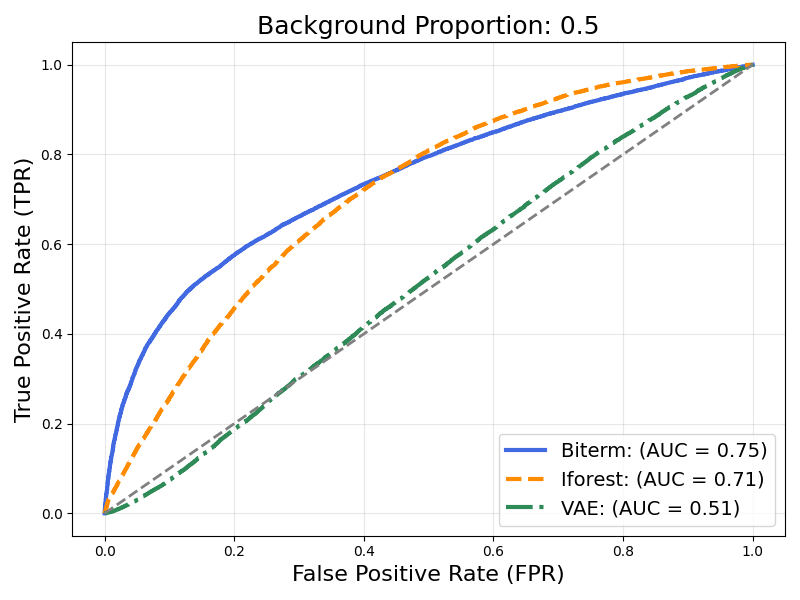}
    \end{minipage}\hfill
    \begin{minipage}{0.48\linewidth}
        \centering
        \includegraphics[width=\linewidth]{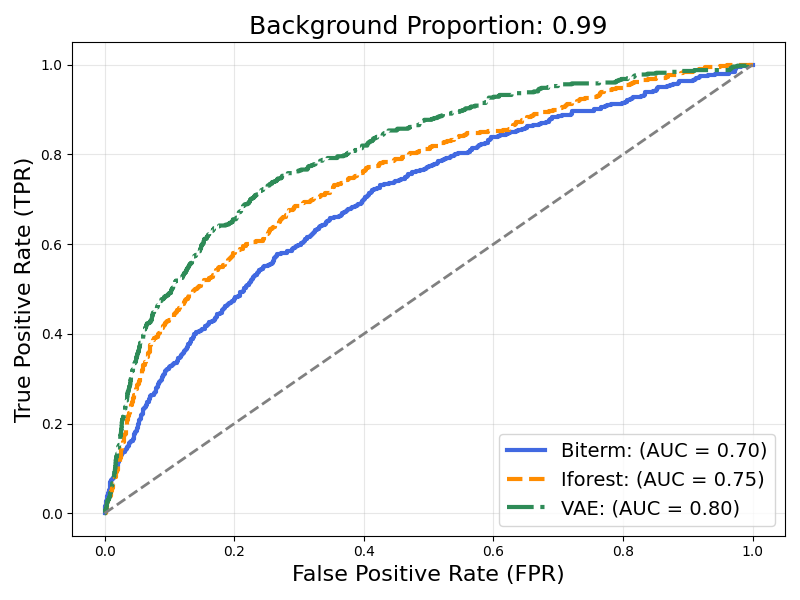}
    \end{minipage}
    \begin{minipage}{0.48\linewidth}
        \centering
        \includegraphics[width=\linewidth]{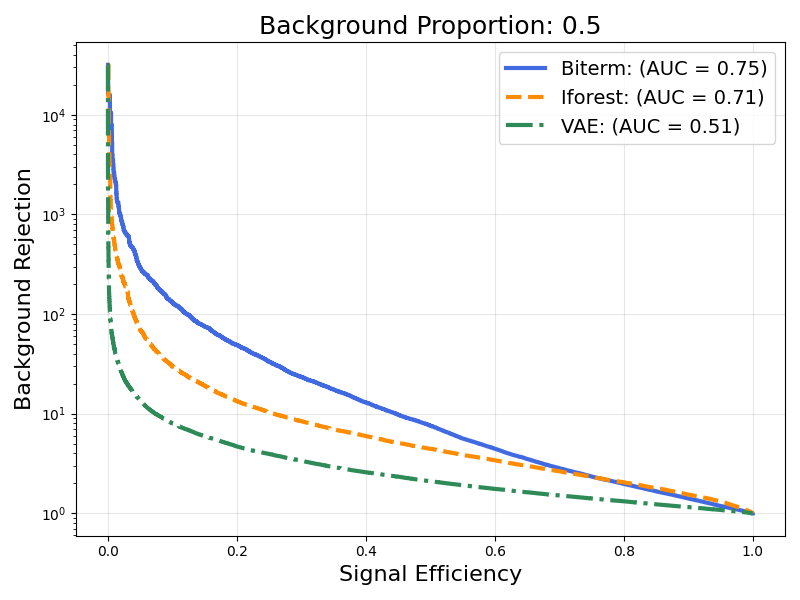}
    \end{minipage}\hfill
    \begin{minipage}{0.48\linewidth}
        \centering
        \includegraphics[width=\linewidth]{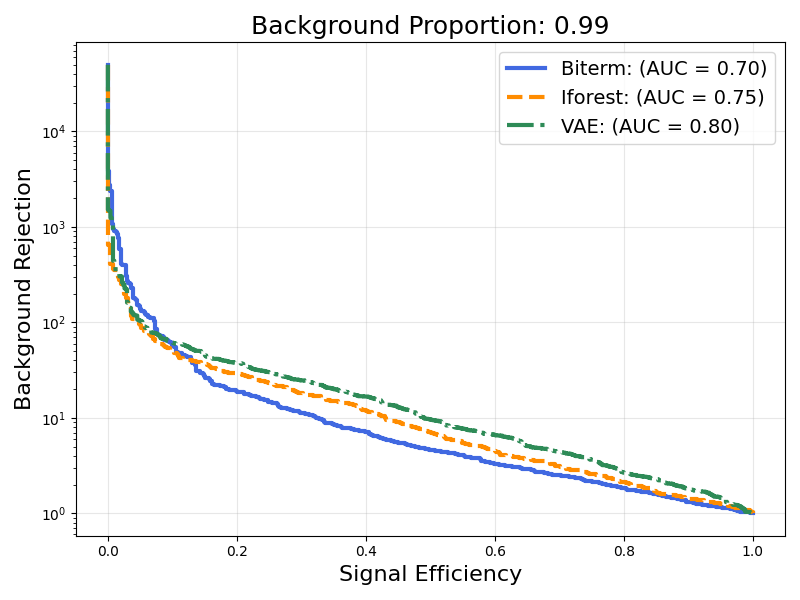}
    \end{minipage}
    \caption{Comparison among the ROC and Background Rejection curves from biterm, isolation forest, and VAE models in double Higgs production with constant coupling $\kappa=+1$ (SM). The left column shows the case of 50\% of background pollution, while the right side shows the 99\% scenario. The upper figures show the ROC curves, which represent the relationship between the true positive rate and false positive rate across many different classification thresholds. The lower figures show the background rejection, which relates the background rejection to the signal efficiency. The legends show the AUC values, which are used to discriminate the ability of each model in disentangling the signal from the background.}
    \label{fig:hh_k+1}
\end{figure}

\begin{figure}[htbp]
    \centering
    \begin{minipage}{0.48\linewidth}
        \centering
        \includegraphics[width=\linewidth]{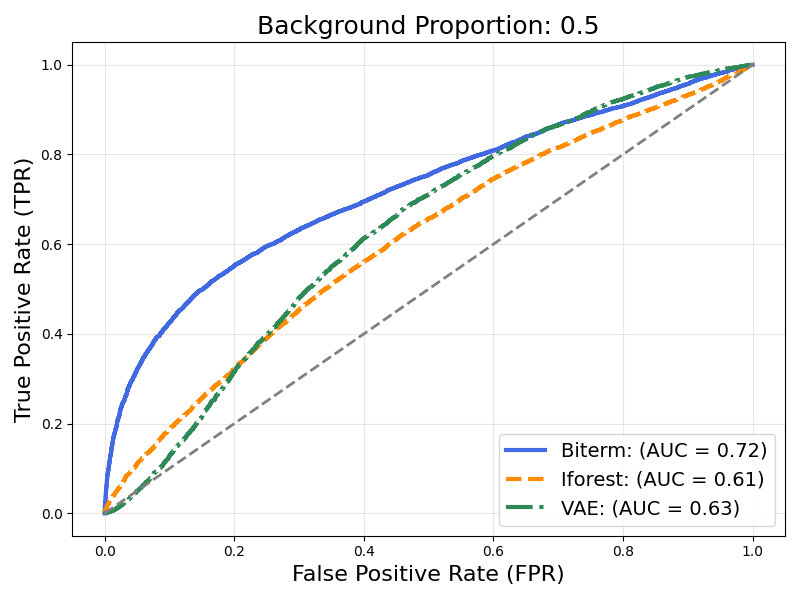}
    \end{minipage}\hfill
    \begin{minipage}{0.48\linewidth}
        \centering
        \includegraphics[width=\linewidth]{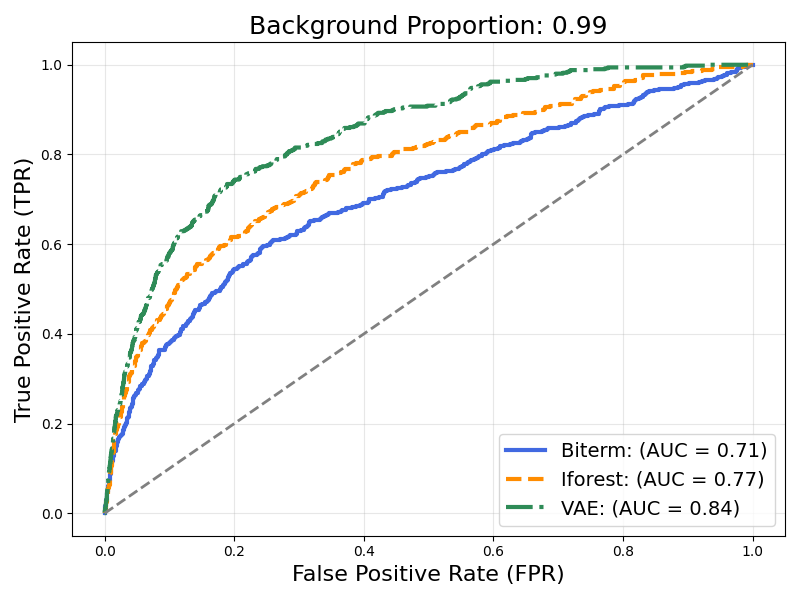}
    \end{minipage}
    \begin{minipage}{0.48\linewidth}
        \centering
        \includegraphics[width=\linewidth]{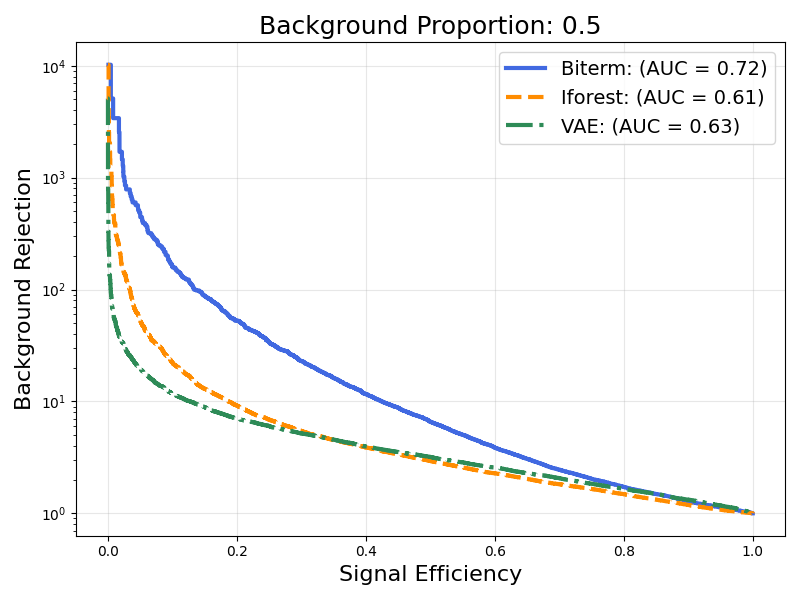}
    \end{minipage}\hfill
    \begin{minipage}{0.48\linewidth}
        \centering
        \includegraphics[width=\linewidth]{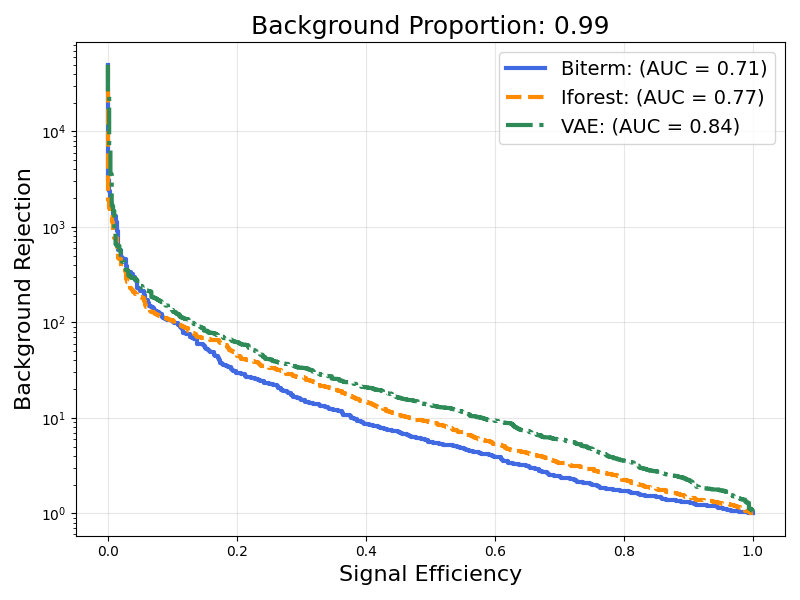}
    \end{minipage}
    \caption{Comparison among the ROC and Background Rejection curves from biterm, isolation forest, and VAE models in double Higgs production with constant coupling $k=+3$ (BSM). The left column shows the case of 50\% of background pollution, while the right side shows the 99\% scenario. The upper figures show the ROC curves, which represent the relationship between the true positive rate and false positive rate across many different classification thresholds. The lower figures show the background rejection, which relates the background rejection to the signal efficiency. The legends show the AUC values, which are used to discriminate the ability of each model in disentangling the signal from the background.}
    \label{fig:hh_k+3}
\end{figure}

\begin{figure}[htbp]
    \centering
    \begin{minipage}{0.48\linewidth}
        \centering
        \includegraphics[width=\linewidth]{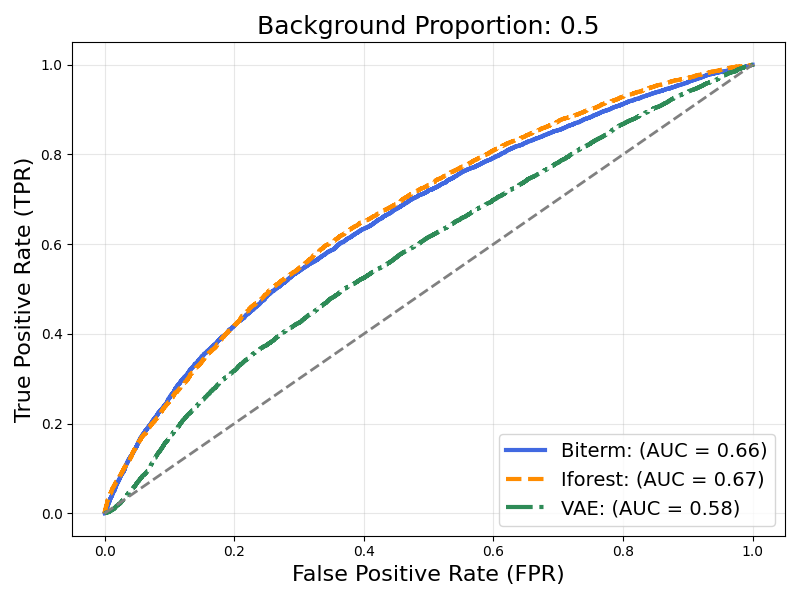}
    \end{minipage}\hfill
    \begin{minipage}{0.48\linewidth}
        \centering
        \includegraphics[width=\linewidth]{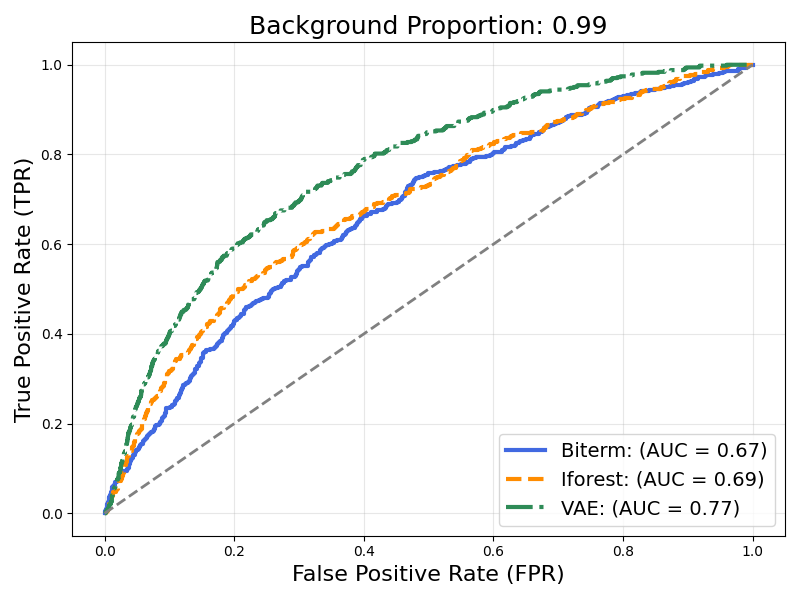}
    \end{minipage}
    \begin{minipage}{0.48\linewidth}
        \centering
        \includegraphics[width=\linewidth]{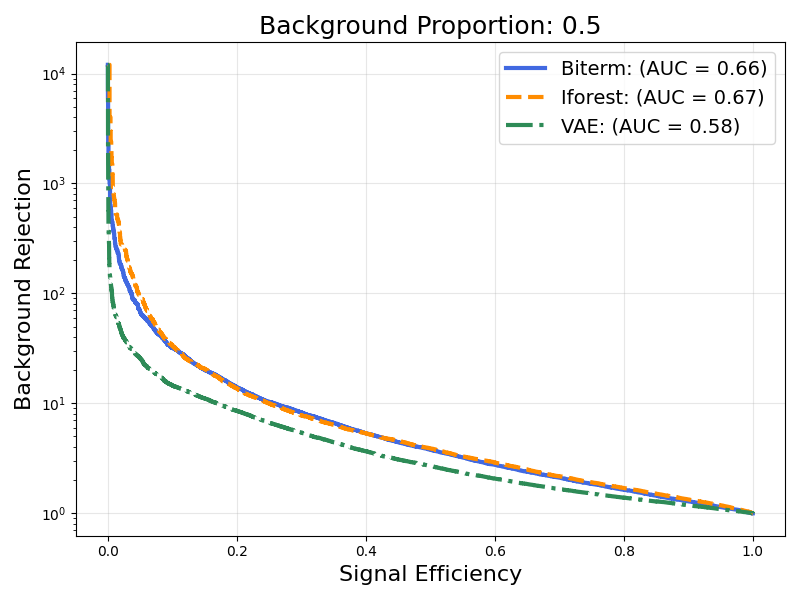}
    \end{minipage}\hfill
    \begin{minipage}{0.48\linewidth}
        \centering
        \includegraphics[width=\linewidth]{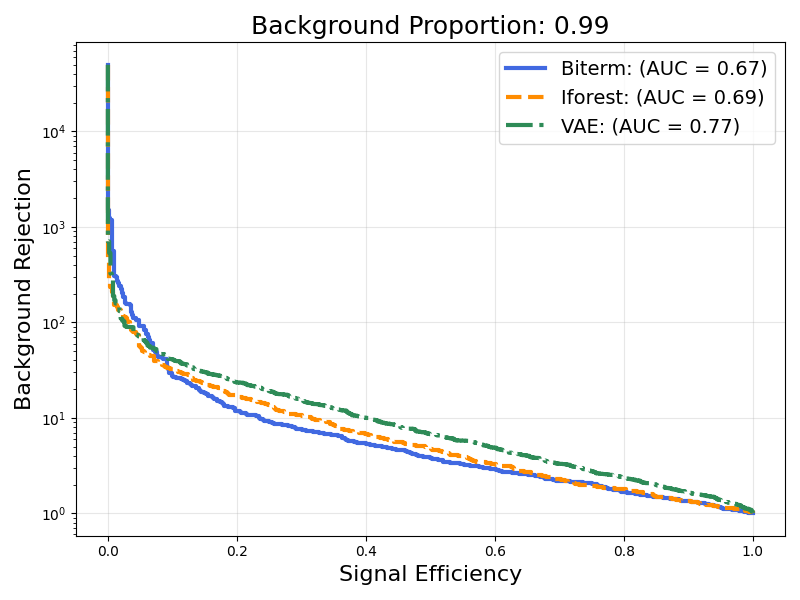}
    \end{minipage}
    \caption{Comparison among the ROC and Background Rejection curves from biterm, isolation forest, and VAE models in double Higgs production with constant coupling $\kappa=+1$ (SM). The left column illustrates the case of 50\% background pollution, while the right side shows the 99\% scenario. The upper figures display the ROC curves, which illustrate the relationship between the true positive rate and false positive rate across various classification thresholds. The lower figures show the background rejection, which relates the background rejection to the signal efficiency. The legends show the AUC values, which are used to discriminate the ability of each model in disentangling the signal from the background.}
    \label{fig:hh_k-1}
\end{figure}

Figure \ref{fig:hh_k+1} shows the SM case ($\kappa=+1$). As in the heavy Higgs cases, the topic modeling model exhibits a similar power of discrimination independently of the background pollution, with $\text{AUC}=0.75$ in the balanced scenario and $\text{AUC}=0.70$ for the extreme case. Figures \ref{fig:hh_k+3} and \ref{fig:hh_k-1} correspond to the BSM $\kappa=+3$ and $\kappa=-1$ cases, respectively, which show a similar behavior to the case $\kappa=+1$, with AUC around $0.7$ in all cases and showing no impact on its performance due to the degree of imbalance of the data.

The IForest and VAE models follow the expected behavior, with a clear improvement when applied to the high pollution background compared to the balanced case. VAE exhibits better separation power with AUC $\sim$ 0.8. However, both the topic model and the outlier detector models demonstrated very competitive results. Table \ref{tab:results_non_resonant_hh} summarizes the AUC results of these cases.

\begin{table}[b!]
\centering
\begin{tabular}{lcc|cc|cc}
\toprule
Model 
& \multicolumn{2}{c}{$\kappa=+1$} 
& \multicolumn{2}{c}{$\kappa=+3$} 
& \multicolumn{2}{c}{$\kappa=-1$} \\

\cmidrule(lr){2-3}
\cmidrule(lr){4-5}
\cmidrule(lr){6-7}

& 50\% bkg & 99\% bkg & 50\% bkg & 99\% bkg & 50\% bkg & 99\% bkg \\
\midrule
BTM     & 0.75 & 0.70 & 0.72 & 0.71 & 0.66 & 0.67 \\
IForest & 0.71 & 0.75 & 0.61 & 0.77 & 0.67 & 0.69 \\
VAE     & 0.51 & 0.80 & 0.63 & 0.84 & 0.58 & 0.77 \\
\bottomrule
\end{tabular}
\caption{
AUC values summarizing the ROC curves for each model and coupling scenario.
Background contamination levels of 50\% and 99\% were considered.
}
\label{tab:results_non_resonant_hh}
\end{table}

\subsection{Effective Operators}

The last case considered in this work involves the insertion of the effective operators $\mathscr{O}_{\tilde{G}}$ and $\mathscr{O}_{qq}^{(1)}$. The set of variables used in the analysis is $\{M,\, p_T\}$ for $\mathscr{O}_{\tilde{G}}$ and $\{\cos\theta^*\}$ for $\mathscr{O}_{qq}^{(1)}$. %, where $f_i$ with $i = \tilde{G},\, qq$ denote the Wilson coefficients of the corresponding dimension-six operators.

\begin{figure}[htbp]
    \centering
    \begin{minipage}{0.48\linewidth}
        \centering
        \includegraphics[width=\linewidth]{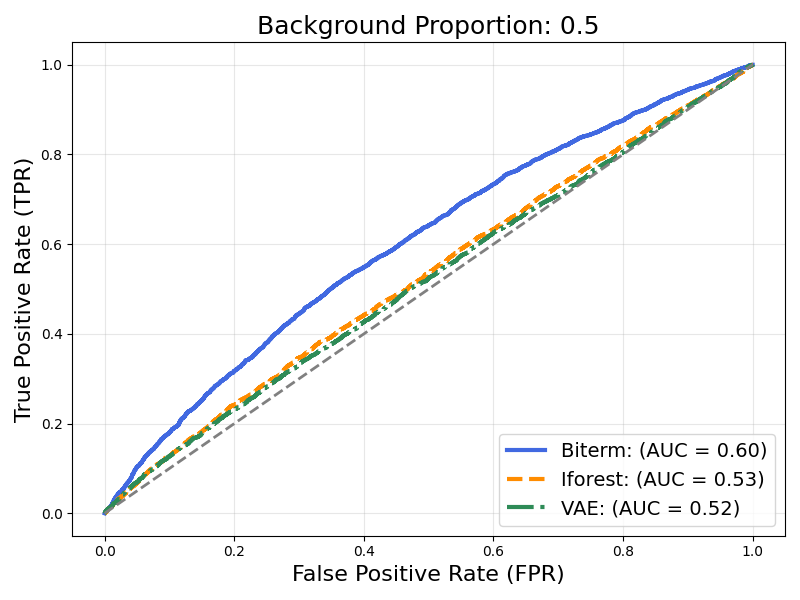}
    \end{minipage}\hfill
    \begin{minipage}{0.48\linewidth}
        \centering
        \includegraphics[width=\linewidth]{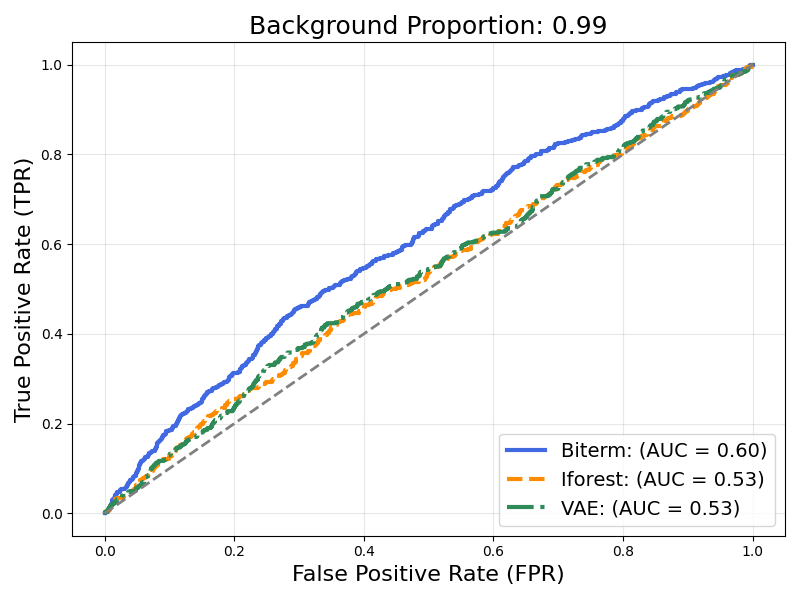}
    \end{minipage}
    \begin{minipage}{0.48\linewidth}
        \centering
        \includegraphics[width=\linewidth]{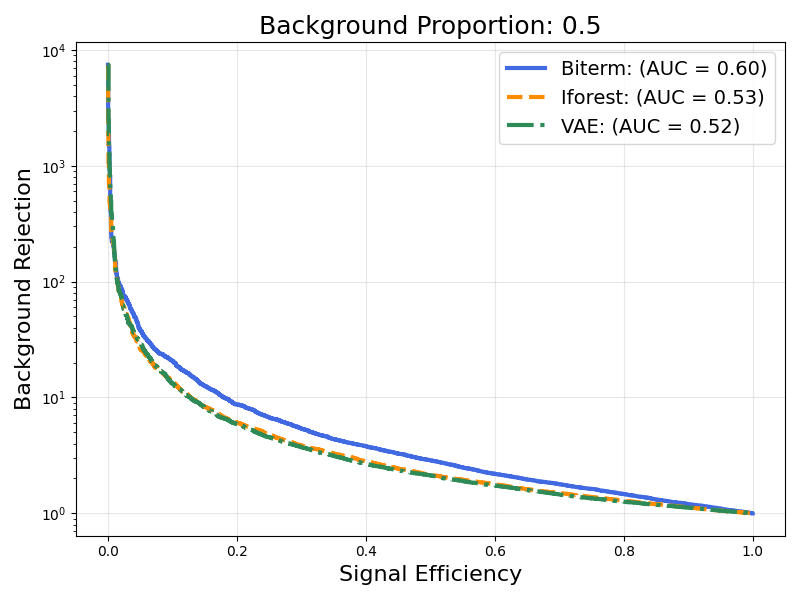}
    \end{minipage}\hfill
    \begin{minipage}{0.48\linewidth}
        \centering
        \includegraphics[width=\linewidth]{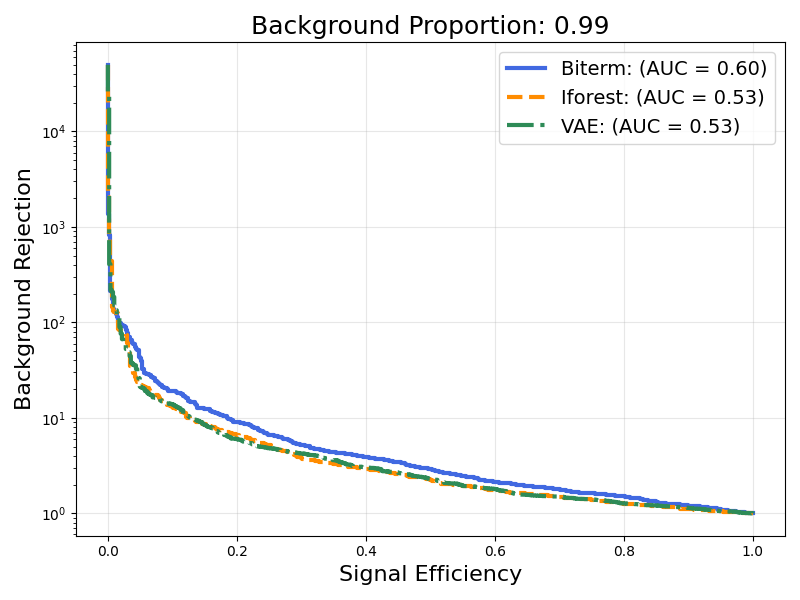}
    \end{minipage}
    \caption{Comparison among the ROC and Background Rejection curves from biterm, isolation forest, and VAE models for the effective operator $\mathscr{O}_{\tilde{G}}$. The left column shows the case of 50\% of background pollution, while the right side shows the 99\% scenario. The upper figures show the ROC curves, which represent the relationship between the true positive rate and false positive rate across many different classification thresholds. The lower figures show the background rejection, which relates the background rejection to the signal efficiency. The legends show the AUC values, which are used to discriminate the ability of each model in disentangling the signal from the background.}
    \label{fig:eff_op_fg}
\end{figure}

\begin{figure}[htbp]
    \centering
    \begin{minipage}{0.48\linewidth}
        \centering
        \includegraphics[width=\linewidth]{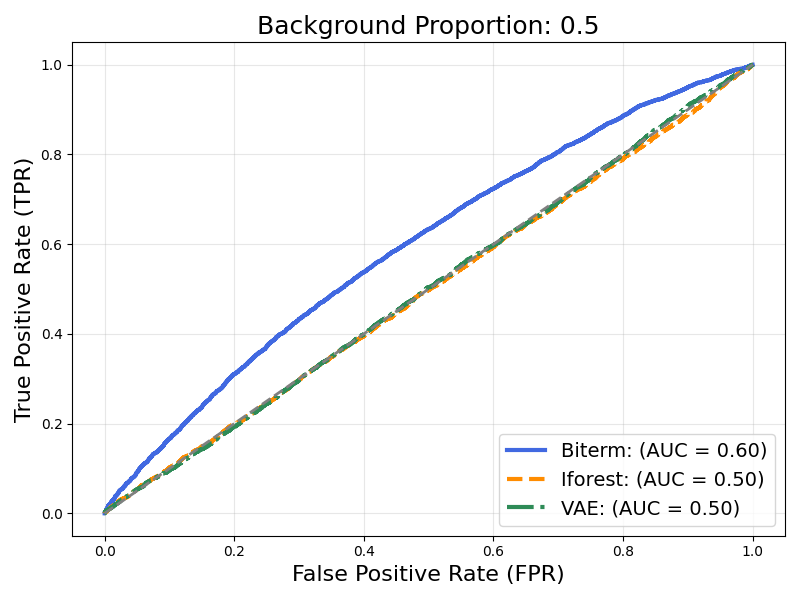}
    \end{minipage}\hfill
    \begin{minipage}{0.48\linewidth}
        \centering
        \includegraphics[width=\linewidth]{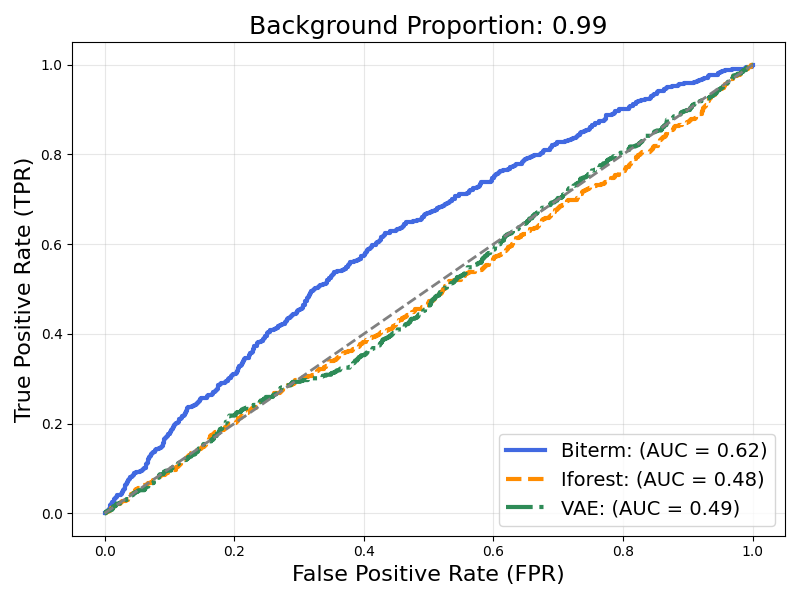}
    \end{minipage}
    \begin{minipage}{0.48\linewidth}
        \centering
        \includegraphics[width=\linewidth]{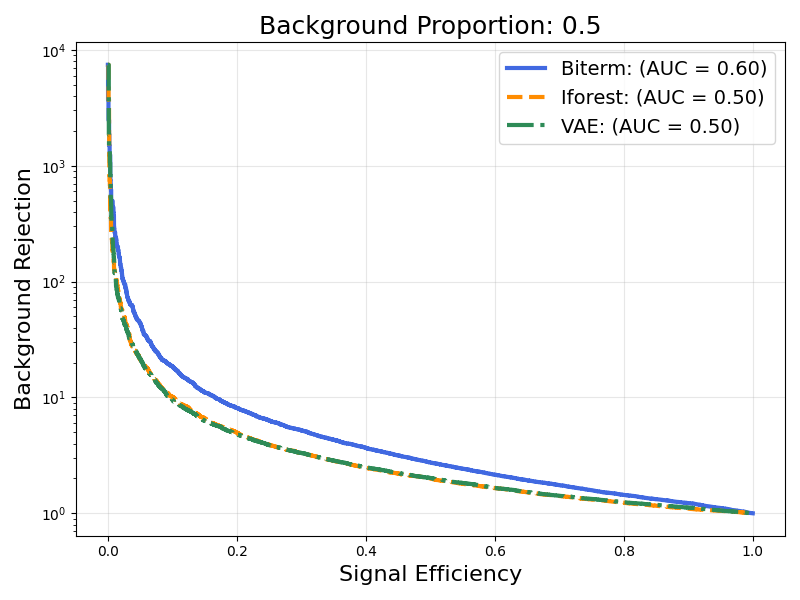}
    \end{minipage}\hfill
    \begin{minipage}{0.48\linewidth}
        \centering
        \includegraphics[width=\linewidth]{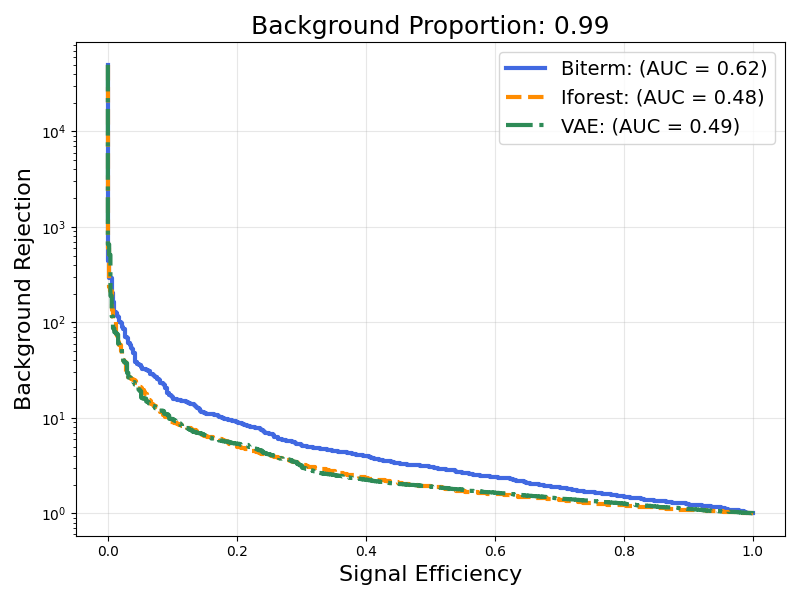}
    \end{minipage}
    \caption{Comparison among the ROC and Background Rejection curves from biterm, isolation forest, and VAE models for the effective operator $\mathscr{O}_{qq}^{(1)}$. The left column illustrates the case of 50\% background pollution, while the right side shows the 99\% scenario. The upper figures display the ROC curves, which illustrate the relationship between the true positive rate and false positive rate across various classification thresholds. The lower figures show the background rejection, which relates the background rejection to the signal efficiency. The legends show the AUC values, which are used to discriminate the ability of each model in disentangling the signal from the background.}
    \label{fig:eff_op_fqq}
\end{figure}

The Figure \ref{fig:eff_op_fg} shows the results for $\mathscr{O}_{\tilde{G}}$. The topic modeling model returns an AUC of 0.6 in both balanced and unbalanced cases. With an AUC $\sim$ 0.5, alternative models are compatible with random guessing, indicating that there is no discriminative power of separation, independently of background pollution. Very similar behavior is found in the case $\mathscr{O}_{qq}^{(1)}$, which is shown in Figure \ref{fig:eff_op_fqq}. In both regimes of background pollution, BTM obtain AUC of around 0.6, while the alternative models perform as a random classifier. Table \ref{tab:results_effective_operators} summarizes these results.

As we can see, effective operators arise as the more challenging case.
In the heavy Higgs cases and in the double Higgs production, it is possible to see a systematic trend in IForest and VAE models to improve your capacity to separate signal from background when moving from 50\% to 99\% background contamination. However, for effective operators, these models do not present any improvement. In contrast to the BTM model, which exhibits a ROC curve consistently above the random-guessing diagonal across the entire threshold range, the observed discriminatory power is stable, although weak, and therefore unlikely to result from statistical fluctuations.

\begin{table}[h]
\centering
\begin{tabular}{lcc|cc}
\toprule
Model 
& \multicolumn{2}{c}{$\mathscr{O}_{\tilde{G}}$}
& \multicolumn{2}{c}{$\mathscr{O}_{qq}^{(1)}$} \\

\cmidrule(lr){2-3}
\cmidrule(lr){4-5}

& 50\% bkg & 99\% bkg & 50\% bkg & 99\% bkg \\
\midrule

BTM     & 0.60 & 0.60 & 0.60 & 0.62 \\
IForest & 0.53 & 0.53 & 0.50 & 0.48 \\
VAE     & 0.52 & 0.53 & 0.50 & 0.49 \\

\bottomrule
\end{tabular}
\caption{
AUC values summarizing the ROC curves for each model and effective operator scenarios.
Background contamination levels of 50\% and 99\% were considered.
}
\label{tab:results_effective_operators}
\end{table}

\section{Discussions}
\label{section:discussions}
We applied the BTM, representing a topic-modeling approach, to three representative scenarios relevant to physics beyond the Standard Model: resonant heavy Higgs boson production decaying into double Higgs and $t\bar{t}$ final states, non-resonant Higgs boson pair production under different trilinear coupling configurations, and scenarios described by effective operators.

A first notable feature of the BTM is its consistent classification performance across all scenarios, largely independent of the degree of class imbalance in the dataset. This is a particularly relevant result, since a strong imbalance is known to significantly degrade the performance of most machine learning algorithms, including highly optimized supervised models.

Although VAE and Isolation Forest are generally regarded as suitable approaches for extreme imbalance regimes, BTM exhibits superior or highly competitive performance in all cases considered. For the resonant scenarios summarized in Table~\ref{tab:results_heavy_higgs}, BTM achieves AUC values above 0.9 even under high background contamination. In contrast, for the $H \to t\bar{t}$ process, none of the anomaly detection models reaches an AUC of 0.9, even in relatively favorable data regimes, highlighting the increased complexity of this final state.

The non-resonant Higgs boson pair production represents a more challenging classification task. In this case, all models exhibit moderate discrimination power, with AUC values ranging from $\sim$ 0.7 to 0.84 in high-background regimes, as shown in Table~\ref{tab:results_non_resonant_hh}. Unlike the resonant heavy Higgs scenarios, where the invariant mass provides a strong discriminative feature, its relevance is significantly reduced in the non-resonant case. Instead, kinematic and angular observables such as $\cos\theta_{WW}$ and $z$ emerge as more informative features, in addition to transverse momentum-related variables ($p_T$ and $k_T$) that are already relevant in the resonant analyses.

The most challenging scenario corresponds to the effective operator cases summarized in Table~\ref{tab:results_effective_operators}. In this regime, BTM displays a weak but consistent classification capability, with AUC values of approximately 0.6, while VAE and Isolation Forest do not exhibit any significant signal discrimination power. Notably, this is the first scenario in which the angular variable $\cos\theta^{*}$ emerges as a relevant observable, capturing subtle patterns that distinguish signal from background. In particular, $\cos \theta^{*}$ appears to be the only variable with discriminative power for the effective operator $\mathscr{O}_{qq}^{(1)}$.

Overall, these results indicate that topic-modeling-based approaches can achieve competitive signal detection performance using exclusively final-state kinematic observables, without the need for jet substructure information, even in regimes characterized by strong background contamination.

\section{Conclusions}
\label{section:conclusions}
In this work, we investigated the potential of topic-modeling-based methods, in particular the Biterm Topic Model, for unsupervised signal detection in LHC collision events using exclusively final-state kinematic observables. The approach was systematically tested across resonant and non-resonant double Higgs production and $t\bar{t}$ production channels via SM and through a heavy new Higgs decay, as well as in scenarios described by effective operators, under both balanced and highly imbalanced background conditions.

Our results demonstrate that BTM provides a robust and stable discrimination capability across all scenarios, with a performance largely insensitive to the degree of background contamination. In resonant heavy Higgs searches, the method achieves AUC values above 0.9 and exhibits excellent background rejection at low signal efficiencies, competing with and, in several cases, surpassing established anomaly-detection models such as VAE and Isolation Forest. For non-resonant Higgs pair production, all models show moderate discrimination power, reflecting the intrinsic complexity of this process, while BTM remains consistently competitive and stable across different constant couplings.

The effective operator scenarios constitute the most challenging regime. In this case, BTM is the only method that maintains a classification performance systematically above random guessing, indicating that it captures subtle latent structures encoded in the final-state observables, even when conventional anomaly-detection models fail to extract meaningful separation.

A key outcome of this study is that topic modeling can achieve competitive signal-background separation without relying on jet substructure observables, using only a compact set of final-state kinematic variables. This highlights the ability of topic models to uncover physically meaningful latent patterns in sparse event representations, offering a complementary strategy to traditional supervised and anomaly-detection approaches.

Although the classification performance in some scenarios remains limited by the intrinsic overlap between signal and background, the stability of the topic-modeling approach under extreme class imbalance and its minimal reliance on feature engineering make it particularly attractive for exploratory searches and model-independent analyses.

Future studies may extend this framework to new sets of observables, approach other physical scenarios, and explore hybrid strategies that combine topic modeling with other techniques. Overall, our results indicate that topic-modeling-based methods constitute a promising and versatile tool for unsupervised new physics searches at the LHC.

\bigskip{}
	
\textbf{Acknowledgments}: This study was financed in part by Conselho Nacional de Desenvolvimento Científico e Tecnológico (CNPq), via Grant No. 307317/2021-8 (A. A.). A. A. also acknowledges support from the FAPESP (No. 2021/01089-1) Grant. 

\bibliography{main.bib}

\end{document}